\documentclass[longauth]{aa} % for the long lists of affiliations 
\usepackage{graphicx}
\usepackage{txfonts}
\begin{document}

\title{The ALPINE-ALMA [C\,{\sc ii}] survey:\\ 
Molecular gas budget in the Early Universe as traced by [C\,{\sc ii}]}

%[CII] insights into the molecular gas in the Early Universe
%molecular gas evolution across the cosmic time
%a first look into the molecular gas content in the Early Universe
%a [C\,{\sc ii}] window on the molecular gas content in the Early Universe

%  \subtitle{I. Overviewing the $\kappa$-mechanism}

\author{M.~Dessauges-Zavadsky\inst{1},
        M.~Ginolfi\inst{1},
        F.~Pozzi\inst{2,3},
        M.~B\'ethermin\inst{4},
        O.~Le F\`evre\inst{4},
        S.~Fujimoto\inst{5,6},
        J.~D.~Silverman\inst{7,8},
        G.~C.~Jones\inst{9,10},
        D.~Schaerer\inst{1},
        A.~L.~Faisst\inst{11},
        Y.~Khusanova\inst{4,12},
        Y.~Fudamoto\inst{1},
        P.~Cassata\inst{13,14},
        F.~Loiacono\inst{2,3},
        P.~L.~Capak\inst{11,5,6},
        L.~Yan\inst{15},
        R. Amorin\inst{16,17},
        S.~Bardelli\inst{3},
        M.~Boquien\inst{18},
        A.~Cimatti\inst{2,19},
        C.~Gruppioni\inst{3},
        N.~P.~Hathi\inst{20},
        E.~Ibar\inst{21},
        A.~M.~Koekemoer\inst{20},
        B.~C.~Lemaux\inst{22},
        D.~Narayanan\inst{23},
        P.~A.~Oesch\inst{1,5},
        G.~Rodighiero\inst{13,14},
        M.~Romano\inst{13,14},
        M.~Talia\inst{2,3},
        S.~Toft\inst{6,5},
        L.~Vallini\inst{24},
        D.~Vergani\inst{3},
        G.~Zamorani\inst{3},
        \and
        E.~Zucca\inst{3}
        }

\institute{
Observatoire de Gen\`eve, Universit\'e de Gen\`eve, 51 Ch. des Maillettes, 1290 Versoix, Switzerland\\
\email{miroslava.dessauges@unige.ch}
\and
Universit\`a di Bologna - Dipartimento di Fisica e Astronomia, Via Gobetti 93/2 - 40129, Bologna, Italy
\and
INAF - Osservatorio di Astrofisica e Scienza dello Spazio di Bologna, via Gobetti 93/3, 40129, Bologna, Italy
\and
Aix Marseille Universit\'e, CNRS, LAM (Laboratoire d’Astrophysique de Marseille) UMR 7326, 13388, Marseille, France
\and
The Cosmic Dawn Center, University of Copenhagen, Vibenshuset, Lyngbyvej 2, 2100 Copenhagen, Denmark
\and
Niels Bohr Institute, University of Copenhagen, Lyngbyvej 2, 2100 Copenhagen, Denmark
\and
Kavli Institute for the Physics and Mathematics of the Universe, The University of Tokyo, Kashiwa, Japan 277-8583 (Kavli IPMU, WPI)
\and
Department of Astronomy, School of Science, The University of Tokyo, 7-3-1 Hongo, Bunkyo, Tokyo 113-0033, Japan
\and
Cavendish Laboratory, University of Cambridge, 19 J. J. Thomson Ave., Cambridge CB3 0HE, UK
\and 
Kavli Institute for Cosmology, University of Cambridge, Madingley Road, Cambridge CB3 0HA, UK
\and
IPAC, M/C 314-6, California Institute of Technology, 1200 East California Boulevard, Pasadena, CA 91125, USA
\and
Max-Planck-Institut f\"ur Astronomie, K\"onigstuhl 17, 69117 Heidelberg, Germany
\and
Dipartimento di Fisica e Astronomia, Universit\`a di Padova, vicolo dell’Osservatorio, 3 35122 Padova, Italy
\and 
INAF - Osservatorio Astronomico di Padova, vicolo dell’Osservatorio 5, 35122 Padova, Italy
\and
The Caltech Optical Observatories, California Institute of Technology, Pasadena, CA 91125, USA
\and
Instituto de Investigaci\'on Multidisciplinar en Ciencia y Tecnolog\'ia, Universidad de La Serena, Ra\'ul Bitr\'an 1305, La Serena, Chile
\and
Departamento de Astronom\'ia, Universidad de La Serena, Av. Juan Cisternas 1200 Norte, La Serena, Chile
\and
Centro de Astronom\'ia (CITEVA), Universidad de Antofagasta, Avenida Angamos 601, Antofagasta, Chile
\and
INAF - Osservatorio Astrofisico di Arcetri, Largo E. Fermi 5, 50125, Firenze, Italy
\and
Space Telescope Science Institute, 3700 San Martin Drive, Baltimore, MD 21218, USA
\and 
Instituto de F\'isica y Astronom\'ia, Universidad de Valpara\'iso, Avda. Gran Breta$\Tilde{\mathrm{n}}$a 1111, Valpara\'iso, Chile
\and 
Department of Physics, University of California, Davis, One Shields Ave., Davis, CA 95616, USA
\and 
Department of Astronomy, University of Florida, 211 Bryant Space Sciences Center, Gainesville, FL 32611 USA
\and
Leiden Observatory, Leiden University, PO Box 9500, 2300 RA Leiden, The Netherlands
}

\date{Received; accepted }

\authorrunning{Dessauges-Zavadsky et~al.}
\titlerunning{Molecular gas budget in the Early Universe as traced by [C\,{\sc ii}]}

% \abstract{}{}{}{}{} 
% 5 {} token are mandatory
  %\abstract
  
\abstract{The molecular gas content of normal galaxies at $z>4$ is poorly constrained, because the commonly used molecular gas tracers become hard to detect at these redshifts. We use the [C\,{\sc ii}] 158~$\mu$m luminosity, recently proposed as a molecular gas tracer, to estimate the molecular gas content in a large sample of main-sequence star-forming galaxies at $z=4.4-5.9$, with a median stellar mass of $10^{9.7}~M_{\odot}$, drawn from the {\it ALMA Large Program to INvestigate [C\,{\sc ii}] at Early times} (ALPINE) survey. The good agreement between molecular gas masses derived from [C\,{\sc ii}] luminosities, dynamical masses, and rest-frame 850~$\mu$m luminosities, extrapolated from the rest-frame 158~$\mu$m continuum, supports [C\,{\sc ii}] as a reliable tracer of molecular gas in our sample. We find a continuous decline of the molecular gas depletion timescale from $z=0$ to $z=5.9$, which reaches a mean value of $(4.6\pm0.8)\times 10^8$~yr at $z\sim 5.5$, only a factor of $2-3$ shorter than in present-day galaxies. This suggests a mild enhancement of star formation efficiency toward high redshifts, unless the molecular gas fraction significantly increases. Our estimates show that the rise in molecular gas fraction as reported previously, flattens off above $z\sim 3.7$ to achieve a mean value of $63\%\pm 3\%$ over $z=4.4-5.9$. This redshift evolution of the gas fraction is in line with the one of the specific star formation rate. We use multi-epoch abundance matching to follow the gas fraction evolution over cosmic time of progenitors of $z=0$ Milky Way-like galaxies in $\sim 10^{13}~M_{\odot}$ halos and of more massive $z=0$ galaxies in $\sim 10^{14}~M_{\odot}$ halos. Interestingly, the former progenitors show a monotonic decrease of the gas fraction with cosmic time, while the latter show a constant gas fraction from $z=5.9$ to $z\sim 2$ and a steep decrease at $z\lesssim 2$. We discuss three possible effects, namely outflows, halt of gas supplying, and over-efficient star formation, which may jointly contribute to the gas fraction plateau of the latter massive galaxies.}

\keywords{galaxies: evolution -- galaxies: high-redshift -- galaxies: ISM -- ISM: molecules}

\maketitle
%
%-----------------------------------------------------------------------

\section{Introduction}

%The cold molecular gas plays a crucial role in galaxy evolution as it provides the fuel for star formation.
%Cold molecular hydrogen, H$_2$, is the fuel for star formation. Therefore, to understand star formation in galaxies and the process by which gas is converted into stars, it is necessary to probe the molecular gas phase of the interstellar medium (ISM). 
Since cold molecular hydrogen, H$_2$, is the fuel for star formation, it is necessary to probe the molecular gas content of galaxies with cosmic time to understand their stellar assembly. With an increasing number of normal star-forming galaxies (SFGs) having measurements of their cold molecular gas mass ($M_{\rm molgas}$), we are starting to bring to light the significant role that molecular gas plays in the evolution of these galaxies, which contribute to about 90\% of the cosmic star formation rate (SFR) density. They are found to follow the star-forming main-sequence (MS), a tight relation between stellar mass ($M_{\rm stars}$) and SFR, which evolves with redshift and has a dispersion of about $\pm 0.3$~dex \citep[e.g.,][]{rodighiero11,speagle14,whitaker14,tasca15,faisst16}. The redshift evolution of the MS is such that, at a given $M_{\rm stars}$, high-redshift galaxies form more stars per unit time than low-redshift galaxies, resulting in an increase of their specific star formation rate (${\rm sSFR = SFR}/M_{\rm stars}$) with redshift. It is now well established that, up to $z\sim 2.5$, the sSFR increase is linked to the observed rise of the molecular gas content of galaxies with redshift \citep[e.g.,][]{saintonge13,genzel15,dessauges17,tacconi18,tacconi20,decarli19}. Likewise, the location of a galaxy in the SFR--$M_{\rm stars}$ plane is primarily governed by its supply (mass) of molecular gas %($\mu_{\rm molgas} = M_{\rm molgas}/M_{\rm stars}$) 
and to some extent also its star formation efficiency (${\rm SFE = SFR}/M_{\rm molgas}$) \citep[e.g.,][]{magdis12,dessauges15,genzel15,silverman15,silverman18,scoville16,tacconi20}.

To explain the high SFR and $M_{\rm molgas}$ of SFGs in the early Universe, it has been proposed that they must be sustained with cold gas accreted from the cosmic web \citep[e.g.,][]{keres05,dekel09}. In this context, the MS may be interpreted in terms of a ``bathtub'' model, in which MS galaxies lie in a quasi-steady state equilibrium whereby star formation is regulated by the available gas reservoir, and whose content is replenished through pristine gas accretion flows and, eventually, diminished by the amount of material galaxies returned to the intergalactic medium through outflows \citep[e.g.,][]{bouche10,dave11,dave12,lilly13,dekel14}. Beside the average growth of SFGs along the MS, simulations suggest SFGs oscillate up and down in sSFR across the MS dispersion, owing to feedback effects altering the gas accretion rates, internal gas transport, and compaction events \citep{tacchella16,orr19}. The bathtub model agrees with most of the scaling relations observed for MS SFGs, such as the Kennicutt-Schmidt star-formation law \citep{kennicutt98b,tacconi13} and the mass-metallicity relation \citep[e.g.,][]{erb06,maiolino08,mannucci10,ginolfihunt19}, and with the dynamically more turbulent galactic disks at high-redshift \citep[e.g.,][]{forster09,wisnioski15,molina17,girard18}. 

While H$_2$ is the most abundant molecule in the Universe, it is nevertheless difficult to detect in cold media, because it features no emission lines with excitation temperatures below 100~K. Fortunately, cold molecular gas is not pure H$_2$, but contains heavier elements like carbon and oxygen, and is mixed with dust grains. Thus, three indirect cold H$_2$ tracers are commonly used to estimate the H$_2$ content of high-redshift galaxies: the CO molecule rotational transitions \citep[][and references therein]{bolatto13}; the dust mass inferred from the fit of the thermal far-infrared (FIR) dust spectral energy distribution (SED) \citep[e.g.,][]{leroy11,magdis11,santini14,bethermin15,kaasinen19}; and the cold dust continuum emission measured in the Rayleigh-Jeans tail regime of the FIR SED \citep[e.g.,][]{scoville14,scoville16,scoville17}. The Plateau de Bure interferometer --~now the Northern Extended Millimeter Array (NOEMA)~-- and the Atacama Large Millimeter/sub-millimeter Array (ALMA) have largely contributed to the census of $M_{\rm molgas}$ in MS SFGs over the peak of the cosmic star formation from $z=0$ to $z\sim 3.5$ \citep[e.g.,][]{daddi10,magnelli12,tacconi13,
tacconi18,saintonge13,saintonge17,santini14,dessauges15,schinnerer16,decarli19,liu19b}. At higher redshifts, both CO and dust become harder to detect, because of (i)~the surface brightness dimming as $(1+z)^4$, (ii)~the lower metallicities expected in distant galaxies making CO dark and dust rare, and (iii)~the ALMA bands only covering high ($J\geq 5$) CO transitions at $z>4.5$, which requires the knowledge of the CO excitation state and gas density to determine the total $M_{\rm molgas}$. Therefore, only two $M_{\rm molgas}$ estimates derived from CO luminosity measurements were reported in MS SFGs at $z>5.5$ to date \citep{dodorico18,pavesi19}. And the dozens of $M_{\rm molgas}$ measurements derived from FIR dust continuum for MS SFGs at $z>4$ \citep{scoville16,liu19b} are largely biased toward massive galaxies with $M_{\rm stars} \gtrsim 10^{11.5}~M_{\odot}$ (and hence high SFRs). 

Clearly, the MS is not yet adequately covered at these high redshifts \citep[see the right panel of Fig.~4 of][]{liu19b}: molecular gas masses of MS SFGs at $z>4$, for a large parameter space of $M_{\rm stars}$ and SFR, still need to be accessed to establish how gas reservoirs and gas consumption timescales change as a function of at least three fundamental parameters, namely the cosmic time, $M_{\rm stars}$, and SFR. The study of the molecular gas content of galaxies over $4<z<6$ is all the more important as such redshift range corresponds to the key evolutionary phase in the early life of galaxies, between their primordial and mature phase, with many fundamental properties of present-day galaxies being established \citep{ribeiro16,feldmann15}. During this early phase, galaxies are known to double their $M_{\rm stars}$ at 5 to 10 times higher rates than at later cosmic times \citep{faisst16,davidzon18}, which may require very efficient star formation and/or considerable supply of molecular gas. 
%However, how may we access to these molecular gas mass measurements?

The C$^+$ radiation, considered as an important coolant of the neutral interstellar medium (ISM), accessible through the [C\,{\sc ii}] line at 158~$\mu$m \citep[one of the strongest line in the FIR spectra; see][]{carilli13} and shown to correlate with the total SFR in galaxies \citep[e.g.,][]{delooze11,delooze14,schaerer19}, has been found to be a good tracer of molecular gas, first at $0.03<z<0.2$ by \citet{hughes17a} and recently over $0<z<6$ by \citet{zanella18}. Such a correlation between [C\,{\sc ii}] luminosity ($L_{\rm CII}$) and $M_{\rm molgas}$ can be exploited to overcome the observational challenge of detecting CO or FIR dust emission in very high-redshift normal SFGs.
%accessible through one of the strongest emission lines in the infrared --~the [C\,{\sc ii}] line at 158~$\mu$m \citep{carilli13}~-- has been empirically found to be a good tracer of molecular gas, first at $0.03<z<0.2$ by \citet{hughes17a} and later over $0<z<6$ by \citet{zanella18}, while for long the [C\,{\sc ii}] luminosity was shown to correlate with the total SFR of galaxies \citep[e.g.,][]{delooze11,delooze14,schaerer19}. This new [C\,{\sc ii}] luminosity--$M_{\rm molgas}$ correlation, if correct, allows us to overcome the observational challenge and cost of detecting CO or FIR dust emission in very high-redshift normal SFGs.
%needed to determine their molecular gas content. 
In this context, our recently completed ALMA Large Program to INvestigate [C\,{\sc ii}] at Early times \citep[ALPINE;][]{lefevre19,bethermin19,faisst19} delivers the first sample of 75 [C\,{\sc ii}] emission detections and 43 upper limits obtained for a representative population of ultraviolet (UV) selected MS SFGs at $z=4.4-5.9$ with ${\rm SFR} \gtrsim 10~M_{\odot}~\rm yr^{-1}$ and $M_{\rm stars} = 10^{8.4}-10^{11}~M_{\odot}$. Relying on the \citet{zanella18} correlation, we use the ALPINE data to provide the first set of molecular gas mass estimates for MS SFGs at $z=4.4-5.9$.

In Sect.~\ref{sect:observations} we summarise the ALPINE survey, the physical properties of galaxies in our survey, and the ALMA observations. Measurements of molecular gas masses obtained using [C\,{\sc ii}] luminosity are presented in Sect.~\ref{sect:molgas-estimates}, together with specific tests of [C\,{\sc ii}] as a reliable molecular gas tracer for the ALPINE galaxies. In Sect.~\ref{sect:compilation} we describe the comparison sample, which includes lower redshift MS SFGs with molecular gas masses determined from CO luminosities. We argue why CO-detected MS galaxies represent a better comparison sample with respect to FIR continuum-detected SFGs having typically large $M_{\rm stars}$. In Sect.~\ref{sect:discussion} we discuss the inferred molecular gas depletion timescales and the molecular gas fractions, which we compare to those of lower redshift CO-detected galaxies. 
%whose molecular gas masses were determined from CO luminosities. 
The evolution of the molecular gas fraction over cosmic time is described in Sect.~\ref{sect:fmolgas-evolution}. We use the multi-epoch abundance matching predictions to connect the progenitors at high redshifts with their descendants at $z=0$. Our main results are summarised in Sect.~\ref{sect:conclusions}. 

Throughout the paper, we assume the $\Lambda$CDM cosmology with $\Omega_m = 0.3$, $\Omega_{\Lambda} = 0.7$ and $H_0 = 70~\rm km~s^{-1}~Mpc^{-1}$, and we adopt the \citet{chabrier03} initial mass function.

%
%-----------------------------------------------------------------------

\section{Observations and physical properties of ALPINE galaxies}
\label{sect:observations}

%The recently completed ALMA Large Program to INvestigate [C\,{\sc ii}] at Early times (ALPINE; Le F\`evre et~al.\ (2019, survey paper), B\'ethermin et~al.\ (2019, data reduction paper), and Faisst et~al.\ (2019, ancillary data paper)) delivers the first representative sample of main-sequence, normal star-forming galaxies at $z=4.4-5.9$ and with star formation rates ${\rm SFR} \gtrsim 10~M_{\odot}~\rm yr^{-1}$ detected in the [C\,{\sc ii}] emission and FIR continuum emission around 158~$\mu$m rest-frame, close to the FIR SED peak. 
The 118 targeted galaxies from the ALPINE survey (\citeauthor{lefevre19} \citeyear{lefevre19} -- survey paper; \citeauthor{bethermin19} \citeyear{bethermin19} -- data reduction paper; \citeauthor{faisst19} \citeyear{faisst19} -- ancillary data paper) are UV-selected galaxies from the COSMic evOlution Survey \citep[COSMOS, 105 galaxies;][]{scoville07} and the Extended Chandra Deep Field South survey
%Great Observatories Origins Deep Survey South (GOODS-S; 
\citep[ECDFS, 13 galaxies;][]{giacconi02}. All galaxies have optical spectroscopy, ensuring reliable rest-frame UV spectroscopic redshift measurements, and benefit from multi-wavelength ground- and space-based imaging from UV to IR. 

The detailed description of the ancillary spectra and photometric data can be found in \citet{faisst19}, together with the redshift measurements and the SED fits. The derived $M_{\rm stars}$ and SFR of ALPINE galaxies are in the range of $M_{\rm stars} = 10^{8.4}-10^{11}~M_{\odot}$ and ${\rm SFR}_{\rm SED} =3-630~M_{\odot}~\rm yr^{-1}$, respectively, following the expected MS at $z\sim 5$. There is a good agreement between ${\rm SFR}_{\rm SED}$ and $\rm SFR_{\rm UV+IR}$, as shown by \citet{schaerer19}. The latter corresponds to the sum of $\rm SFR_{UV}$, measured from the UV luminosity at 1500~\AA\ rest-frame (uncorrected for dust attenuation), and $\rm SFR_{IR}$, measured from the rest-frame 158~$\mu$m dust continuum emission flux and the FIR SED template of \citet{bethermin17} to infer the total IR luminosity ($L_{\rm IR}$) integrated between 8~$\mu$m and 1000~$\mu$m as described in \citet{bethermin19}. Throughout the paper, we adopt $M_{\rm stars}$ listed in Table~A1 of \citet{faisst19}, based on photometry that includes the {\it Spitzer} IR imaging, and $\rm SFR_{UV+IR}$ derived from the UV magnitudes listed in Table~A1 of \citet{faisst19} and $L_{\rm IR}$ given in Table~B1 of \citet{bethermin19}. For galaxies undetected in the FIR dust continuum (95 ALPINE galaxies), we consider only $\rm SFR_{UV}$ throughout the paper. 
%assuming that $\rm SFR_{IR}$ is negligible. 
\citet{schaerer19} discuss in detail the possible amount of $\rm SFR_{IR}$, the dust-obscured star formation rate, in these 95 ALPINE galaxies and find that their total SFR can be underestimated by a factor of 1.6, on average, according to the average empirically-calibrated relation between infrared excess (${\rm IRX} = L_{\rm IR}/L_{\rm UV}$) and UV spectral slope ($\beta$; $f_{\lambda} \propto \lambda^{\beta}$), derived by \citet{fudamoto20} for the ALPINE sample from median stacking of individual continuum images in bins of $\beta$. For the majority of the 95 ALPINE galaxies, however, $\rm SFR_{IR}$ turns out to be small ($\lesssim 40\%$ of $\rm SFR_{UV}$), since their UV spectral slope is fairly blue. We would like to foreshadow that none of our conclusions change when we take the possible underestimation of the total SFR into account.

The ALMA observations were carried out in band~7 during Cycles~5 and 6, and completed in February 2019. Band~7 ($275-373$~GHz) covers the [C\,{\sc ii}]~158~$\mu$m line from $z=4.1$ to $z=5.9$, but to avoid an atmospheric absorption no source was included in the redshift range of $z=4.6-5.1$. Each target was observed for $15-25$ minutes of on-source time, with the phase center positioned at the rest-frame UV position of the target and one spectral window in the lower-frequency sideband tuned to the [C\,{\sc ii}] frequency redshifted by the rest-frame UV spectroscopic redshift of that target \citep{faisst19}. The other three spectral windows were used for the FIR continuum around 158~$\mu$m rest-frame, close to the FIR SED peak. The ALMA visibility calibration, cleaning, and imaging were performed using the Common Astronomy Software Applications package \citep[CASA;][]{mcmullin07}, as described in detail in \citet{bethermin19}. The resulting root-mean-square noise (RMS) of the 118 [C\,{\sc ii}] data cubes ranges between $0.2~\rm mJy~beam^{-1}$ and $0.55~\rm mJy~beam^{-1}$ per $25~\rm km~s^{-1}$ channel for an angular resolution varying between $0.72\arcsec$ (minimum minor axis) and $1.6\arcsec$ (maximum major axis). The continuum sensitivity varies with frequency. We reach a mean RMS of $50~\mu\rm Jy~beam^{-1}$ for ALPINE galaxies at $z=4.4-4.6$, and $28~\mu\rm Jy~beam^{-1}$ for ALPINE galaxies at $z=5.1-5.9$. The ALMA dataset leads to robust [C\,{\sc ii}] emission detections for 75 ALPINE galaxies and robust FIR dust continuum emission detections for 23 ALPINE galaxies, with a signal-to-noise ratio (SNR) larger than 3.5 corresponding to 95\% purity threshold of both the [C\,{\sc ii}] line and FIR continuum \citep{bethermin19}. Throughout the paper, we consider the $2\,\sigma$-clipped [C\,{\sc ii}] fluxes\footnote{The $2\,\sigma$-clipped flux corresponds to the flux integrated within the region around the source defined by the contour level at $\rm SNR = 2$ in the moment-zero map. The $2\,\sigma$-clipped fluxes are similar to the 2D-fit fluxes obtained from two-dimensional elliptical Gaussian fits over a $3''$ fitting box around the source \citep[see Fig.~16 in][]{bethermin19}.}, and the FIR continuum fluxes derived using the 2D elliptical Gaussian fits. For the 43 ALPINE targets with no [C\,{\sc ii}] detections, we consider the ``secure'' $3\,\sigma$ upper limits\footnote{The ``secure'' $3\,\sigma$ upper limits on [C\,{\sc ii}] fluxes are calculated by adding the $3\,\sigma$ RMS of the noise to the highest flux measured in $1''$ around the phase center in visibility-tapered velocity-integrated flux maps \citep{bethermin19}.} on [C\,{\sc ii}] fluxes listed in Table~C2 of \citet{bethermin19}.

At the achieved angular resolutions, with an average circularized beam of $0.9\arcsec$, corresponding to $\sim 5.3-6.1$~kpc at $z = 4.4-5.9$, about 2/3 of the ALPINE [C\,{\sc ii}]-detected galaxies are moderately spatially resolved in the [C\,{\sc ii}] velocity-integrated intensity maps \citep{bethermin19,lefevre19,fujimoto19}, 
%and velocity field maps \citep{jones19}, 
meaning their intrinsic (total) sizes as seen in [C\,{\sc ii}] emission are about the size of the beam, or a significant fraction thereof, as illustrated by the spectacular object studied by \citet{jones19}. A large diversity of [C\,{\sc ii}] emission morphologies is observed, from compact/unresolved objects, objects appearing as very extended \citep{fujimoto19,ginolfi20}, to objects showing double, or more, merger-like components \citep{jones19}. From our morpho-kinematic visual classification, described in \citet{lefevre19}, based on the [C\,{\sc ii}] emission and velocity field and multi-band optical to IR images, we find signatures of possibly interacting systems for 31 ALPINE [C\,{\sc ii}]-detected galaxies, while only 9 ALPINE galaxies are likely rotation-dominated, indicating that the mass assembly through merging process is frequent at these redshifts for MS SFGs. In what follows, we exclude the 31 galaxies classified as mergers to work with a sample of galaxies where robust measurements of their physical properties can be determined, since deblending the [C\,{\sc ii}] and dust continuum emissions in closely interacting multi-component systems is difficult with the currently available ALMA data \citep{bethermin19}. Therefore, our final sample consists of 87 ALPINE galaxies, of which 44 are detected in [C\,{\sc ii}], while 43 only have [C\,{\sc ii}] upper limits.

%
%-----------------------------------------------------------------------

\section{Molecular gas mass estimates}
\label{sect:molgas-estimates}

\subsection{[C\,{\sc ii}] as a tracer of cold molecular gas}

%The molecular hydrogen, H$_2$, may be the most abundant molecule in the Universe, cold H$_2$ with temperatures below 100~K is no less invisible in emission. Fortunately, molecular gas is not pure H$_2$ and contains heavier elements like carbon and oxygen, as well as dust grains. Therefore, different indirect H$_2$ tracers are used to study the molecular gas content of galaxies, with the most popular being the molecular CO emission \citep{bolatto13} and the thermal dust continuum emission \citep{leroy11,magdis11,scoville14}. However, both CO and dust may become hard to access at very high redshift, because of the lower metallicities expected in distant galaxies making CO dark and dust rare, and because of the ALMA bands allowing to cover only high $J\geq 5$ CO transitions at $z>4.5$. 
\citet{zanella18} have recently proposed the use of [C\,{\sc ii}] emission as a tracer of molecular gas by finding a tight empirical correlation, with a 0.3~dex dispersion, between the [C\,{\sc ii}] luminosity and molecular gas mass derived using mainly the CO tracer \citep[see also][]{hughes17a}. This relation seems to hold regardless of the MS or starburst nature of galaxies, redshift (from $z=0$ to $z=6$), and metallicity (from $12+\log({\rm O/H}) = 7.9$ to $12+\log({\rm O/H}) = 8.8$). \citet{zanella18} motivate their finding with an explanation of the $L_{\rm CII}/L_{\rm IR}$ deficit observed in ultra-luminous IR galaxies (ULIRGs) and high-redshift starbursts. Indeed, if $L_{\rm CII}$ traces $M_{\rm molgas}$ (and $L_{\rm IR}$ the SFR), then the [C\,{\sc ii}] deficit reflects shorter molecular gas depletion timescales in ULIRGs and distant starbursts, consistently with measurements by \citet{daddi10} and \citet{genzel10}.

From the theoretical point of view, the origin of the emission of [C\,{\sc ii}] is complex, because different ISM phases --~ionised, neutral, and molecular~-- are contributing to it. As a result, one needs to establish whether the fraction of [C\,{\sc ii}] emission arising from photodissociation regions \citep[PDRs;][]{stacey91,malhotra01,cormier15,diaz17}, produced by the UV radiation from hot stars heating the outer layers of molecular clouds and associated with both the interface layer of neutral gas as well as ionised gas in the H\,{\sc ii} region itself, is dominating (or not dominating) that arising from the CO photodissociation into C and C$^+$ in the cold neutral medium of molecular clouds \citep{maloney88,madden93,wolfire10,narayanan17}. In the PDR case C$^+$ is rather tracing star formation, while in the CO photodissociation case C$^+$ emission emerges from the molecular phase. 

The [C\,{\sc ii}] line has raised considerable interest in galaxies at $z\gtrsim 5$, leading several numerical simulations to model [C\,{\sc ii}] and to suggest that its emission is dominated at level of $>60-85$\% by molecular clouds more than by diffuse ionised gas \citep{vallini15,pallottini17,accurso17,olsen18}. Indeed, CO and [C\,{\sc ii}] emission maps of the high-redshift galaxy simulated by \citet{vallini18} and \citet{pallottini17}, respectively, clearly show the same morphology with similar spatial distributions on 30~pc scale. In our Milky Way, dense PDRs and CO-dark H$_2$ gas are also the dominant [C\,{\sc ii}] emitters, responsible for $\sim 55$\% of the total [C\,{\sc ii}] emission, while the diffuse ionised gas and diffuse neutral gas contribute to $\sim 20$\% and $\sim 25$\%, respectively (\citeauthor{pineda14} \citeyear{pineda14}; and see also the simulation predictions from \citeauthor{li18} \citeyear{li18}). As \citet{zanella18} warn, when using [C\,{\sc ii}] as a molecular gas tracer one needs to be aware that since the C$^+$ emission might not fully emerge from one single gas phase, the measured [C\,{\sc ii}] luminosity might overestimate the luminosity arising from the molecular gas even in high-redshift galaxies. On the other hand, as C$^+$ is emitted only in regions where star formation is taking place, the molecular gas not illuminated by stars would not be detected.

%
%-----------------------------------------------------------------------
\begin{figure}
\centering
\includegraphics[width=8.5cm,clip]{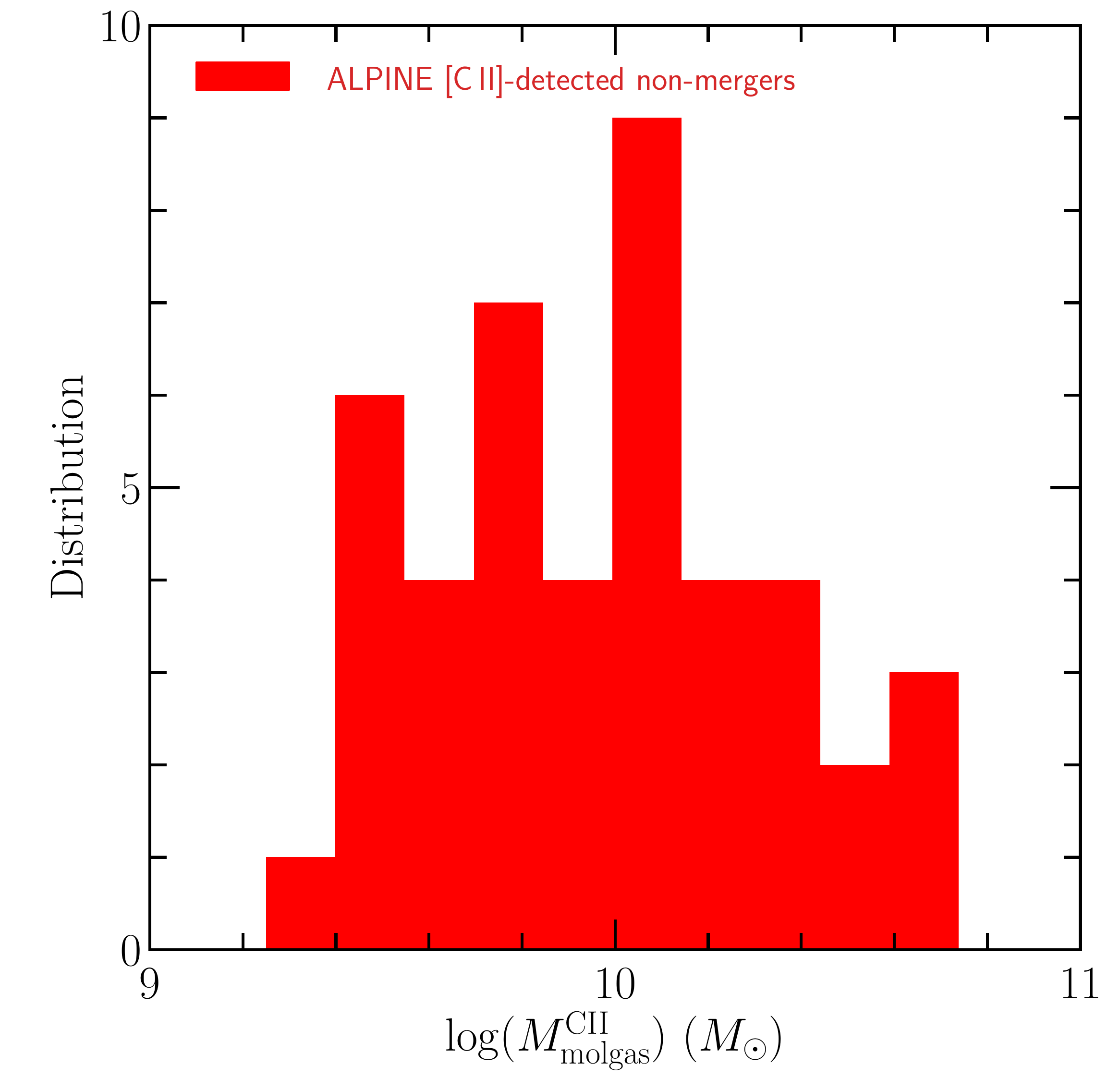}
\caption{Distribution of molecular gas masses of the 44 ALPINE [C\,{\sc ii}]-detected non-merger galaxies at $z=4.4-5.9$. The molecular gas masses are derived using the calibration of \citet{zanella18} between [C\,{\sc ii}] luminosity and molecular gas mass (Eq.~(\ref{eq:zanella})).}
\label{fig:Mmolgas-distributions}
\end{figure}
%
%-----------------------------------------------------------------------

Applying the calibration of \citet{zanella18} between [C\,{\sc ii}] luminosity and molecular gas mass:
\begin{equation}
\label{eq:zanella}
\log \left(\frac{L_{\rm CII}}{L_{\odot}}\right) = (-1.28\pm 0.21) + (0.98\pm 0.02) \log \left(\frac{M^{\rm CII}_{\rm molgas}}{M_{\odot}}\right)
\end{equation}
to the 44 ALPINE [C\,{\sc ii}]-detected non-merger galaxies with $\log(L_{\rm CII}/L_{\odot}) = 7.8-9.2$, in the regime tested by \citet{zanella18}, we obtain molecular gas masses in the range of $\log(M^{\rm CII}_{\rm molgas}/M_{\odot}) = 9.2-10.8$ for these MS SFGs at $z=4.4-5.9$, as shown by the $M^{\rm CII}_{\rm molgas}$ distribution in Fig.~\ref{fig:Mmolgas-distributions}. We calculate the error bars on the [C\,{\sc ii}]-estimated molecular gas masses by summing in quadrature the relative uncertainty of [C\,{\sc ii}] fluxes \citep[see][]{bethermin19} and the 0.3~dex dispersion of the $L_{\rm CII}$--$M^{\rm CII}_{\rm molgas}$ calibration \citep{zanella18}.

%
%-----------------------------------------------------------------------
\begin{figure}
\centering
\includegraphics[width=8.5cm,clip]{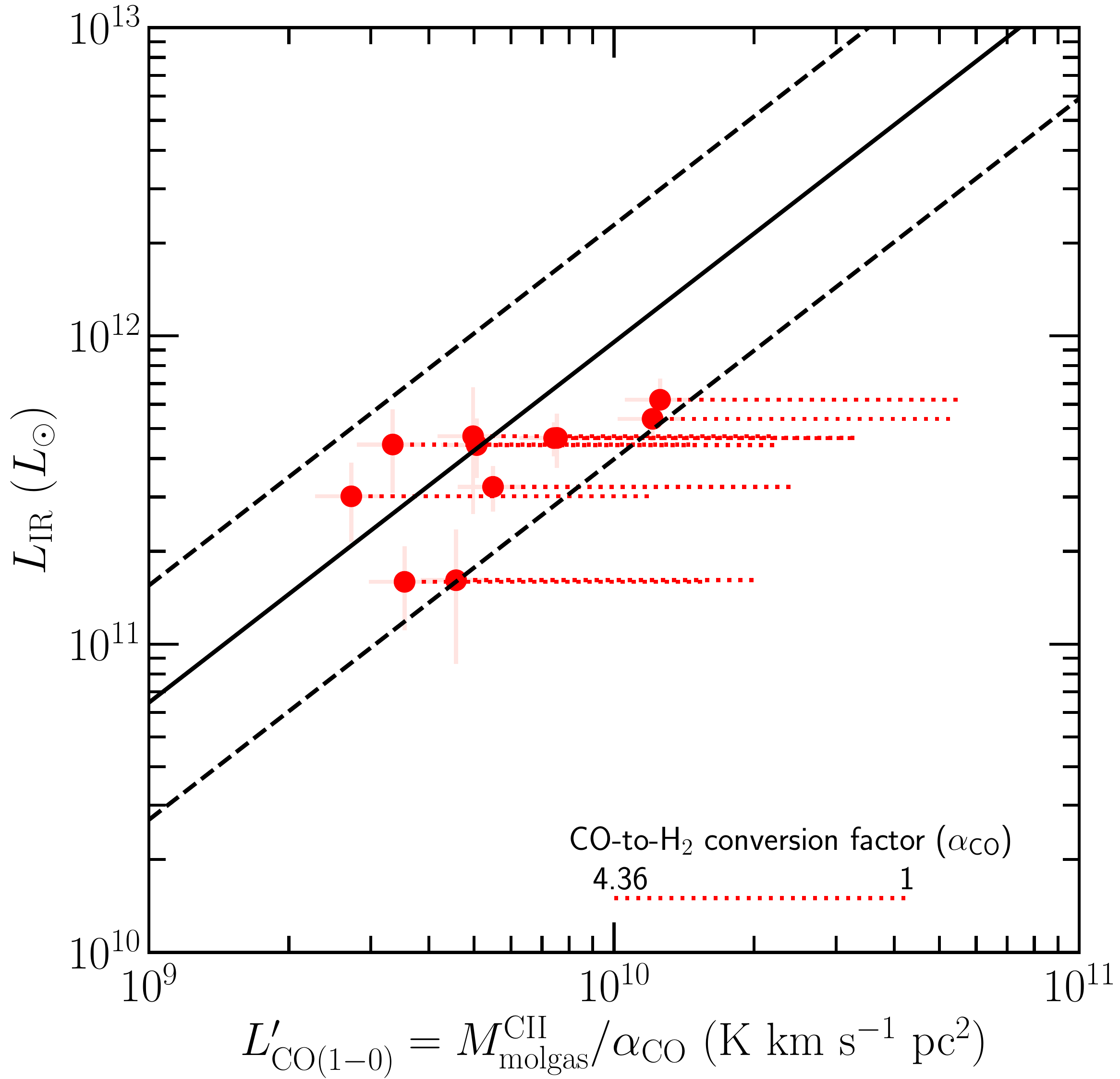}
\caption{IR luminosities measured for 11 ALPINE FIR continuum-detected non-merger galaxies \citep{bethermin19} as a function of their CO(1--0) luminosities inferred from the [C\,{\sc ii}] molecular gas masses and a range of CO-to-H$_2$ conversion factors (dotted red lines) from the Milky Way value of $4.36~M_{\odot}~(\rm K~km~s^{-1}~pc^2)^{-1}$ (on the left) to the starburst value of $1~M_{\odot}~(\rm K~km~s^{-1}~pc^2)^{-1}$ (on the right). The solid black line shows the best-fit of \citet{dessauges15} of the empirical $L_{\rm IR}$--$L'_{\rm CO(1-0)}$ relation with the $1\,\sigma$ dispersion of 0.38~dex (dashed black lines). Within this dispersion the ALPINE [C\,{\sc ii}]-derived molecular gas masses lie on the relation for the Milky Way $\alpha_{\rm CO}$.}
\label{fig:LIR-LCO}
\end{figure}
%
%-----------------------------------------------------------------------

\subsection{Other cold molecular gas mass tracers}
\label{sect:Mmolgas}

In what follows, for a subset of the ALPINE sample we cross-correlate the [C\,{\sc ii}]-derived molecular gas mass estimates with molecular gas masses inferred using other molecular gas tracers to check the robustness of [C\,{\sc ii}] as the tracer of cold molecular gas in our sample of $4.4<z<5.9$ MS SFGs.

\subsubsection{The IR versus CO luminosity relation}

We can use the well established empirical relation between IR luminosity and CO(1--0) luminosity measurements \citep{daddi10,carilli13,sargent14,dessauges15} 
%illustrative of the star formation law, 
to test if the derived $M^{\rm CII}_{\rm molgas}$ agree with the measured $L_{\rm IR}$ along the expected relation. This relation, which spans almost 5 orders of magnitude in $L_{\rm IR}$ from $10^9~L_{\odot}$ to $10^{13.5}~L_{\odot}$, was found to be valid for a variety of galaxy types from MS galaxies, starbursts to mergers at redshifts between $z=0$ and $z\sim 5.3$. In Fig.~\ref{fig:LIR-LCO} we show the IR luminosities measured for 11 ALPINE [C\,{\sc ii}]-detected non-merger galaxies as a function of the CO(1--0) luminosities inferred from the [C\,{\sc ii}] molecular gas masses and a range of CO-to-H$_2$ conversion factors ($\alpha_{\rm CO}$) from the Milky Way value of $4.36~M_{\odot}~\rm (K~km~s^{-1}~pc^2)^{-1}$ to the starburst value of $1~M_{\odot}~\rm (K~km~s^{-1}~pc^2)^{-1}$ \citep{bolatto13}. We find that for the Milky Way CO-to-H$_2$ conversion factor all ALPINE galaxies fall within the 0.38~dex dispersion of the IR luminosity versus CO(1--0) luminosity relation, $\log(L_{\rm IR}/L_{\odot}) = (1.17\pm 0.03) \log(L'_{\rm CO(1-0)}/L_{\odot}) + (0.28\pm 0.23)$, calibrated by \citet{dessauges15}, but comparable to \citet{carilli13}. 

%
%-----------------------------------------------------------------------

\subsubsection{The dust continuum molecular gas masses}
\label{sect:Mmolgas-850um}

At long-wavelengths in the Rayleigh-Jeans tail regime ($\lambda_{\rm rest} \gtrsim 250$~$\mu$m), the thermal dust emission is optically thin and the observed flux density is directly dependent on the mass of dust, the dust opacity coefficient, and the mean temperature of dust contributing to the emission at these wavelengths \citep{scoville16}. By assuming that the molecular gas dominates the overall gas budget (the atomic and ionised gas content being negligible) and fixing a dust-to-gas mass ratio, we can then recover the molecular gas mass from the derived dust mass. The rest-frame 850~$\mu$m luminosity ($L_{850\mu \rm m}$) was found to exhibit a tight correlation with the ISM molecular gas mass and is now frequently used as a molecular gas tracer \citep{scoville14,scoville16,scoville17,hughes17b,privon18,kaasinen19}. The difficulty remains in deriving $L_{850\mu \rm m}$ from often a single-band FIR continuum measurement, since this requires us to assume a dust opacity coefficient and a mean dust temperature, or to know the FIR SED characteristic of the studied galaxies.

%
%-----------------------------------------------------------------------
\begin{figure}
\centering
\includegraphics[width=8.5cm,clip]{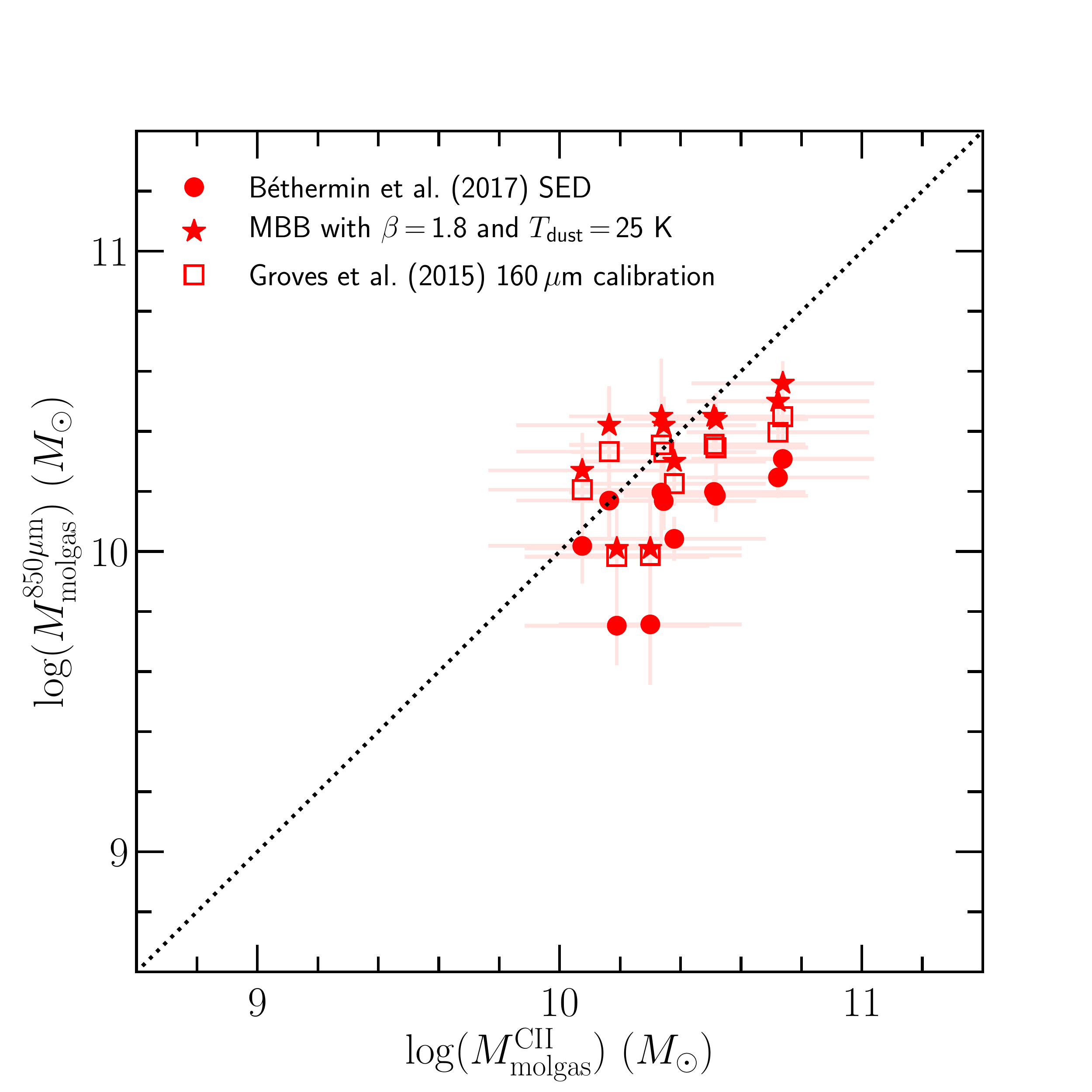}
\caption{Comparison of molecular gas masses of the 11 ALPINE FIR continuum-detected non-merger galaxies as derived from the [C\,{\sc ii}] luminosity (Eq.~(\ref{eq:zanella})) and the rest-frame 850~$\mu$m luminosity (Eq.~(\ref{eq:Scoville})). The monochromatic rest-frame 850~$\mu$m luminosity is extrapolated from the measured rest-frame 158~$\mu$m luminosity by assuming either the FIR SED template of \citet{bethermin17} (filled circles), or the MBB curve with $\beta=1.8$ and $T_{\rm dust} = 25$~K as adopted by \citet{scoville16,scoville17} (filled stars). The open squares show the molecular gas masses derived directly from the measured rest-frame 158~$\mu$m luminosity using the calibration of \citet{groves15}, obtained for local galaxies, between {\it Herschel} PACS 160~$\mu$m monochromatic luminosity and gas mass. The dotted line is the one-to-one relation. Overall, there is a good agreement between $M^{\rm CII}_{\rm molgas}$ and the different molecular gas masses estimated from the rest-frame 158~$\mu$m dust continuum luminosity, even though an average overestimate of 0.3~dex is found when considering the \citet{bethermin19} SED (see text for details).} 
%The filled circles show the comparison of molecular gas masses of the 11 non-merger ALPINE FIR continuum-detected galaxies as derived from the [C\,{\sc ii}] luminosities (Eq.~(\ref{eq:zanella})) and the rest-frame 850~$\mu$m luminosities (Eq.~(\ref{eq:Scoville})). The monochromatic rest-frame 850~$\mu$m luminosity is extrapolated from the measured rest-frame 158~$\mu$m luminosity by assuming the FIR SED template of \citet{bethermin17}, or the modified blackbody curve with $\beta=1.8$ and $T_{\rm dust} = 25$~K as adopted by \citet{scoville16,scoville17} (filled stars). The open squares show the molecular gas masses derived directly from the measured rest-frame 158~$\mu$m luminosity using the calibration of \citet{groves15}, obtained for local galaxies, between {\it Herschel} PACS 160~$\mu$m monochromatic luminosity and gas mass.
\label{fig:Mmolgas-Scoville}
\end{figure}
%
%-----------------------------------------------------------------------
\begin{figure}
\centering
\includegraphics[width=8.5cm,clip]{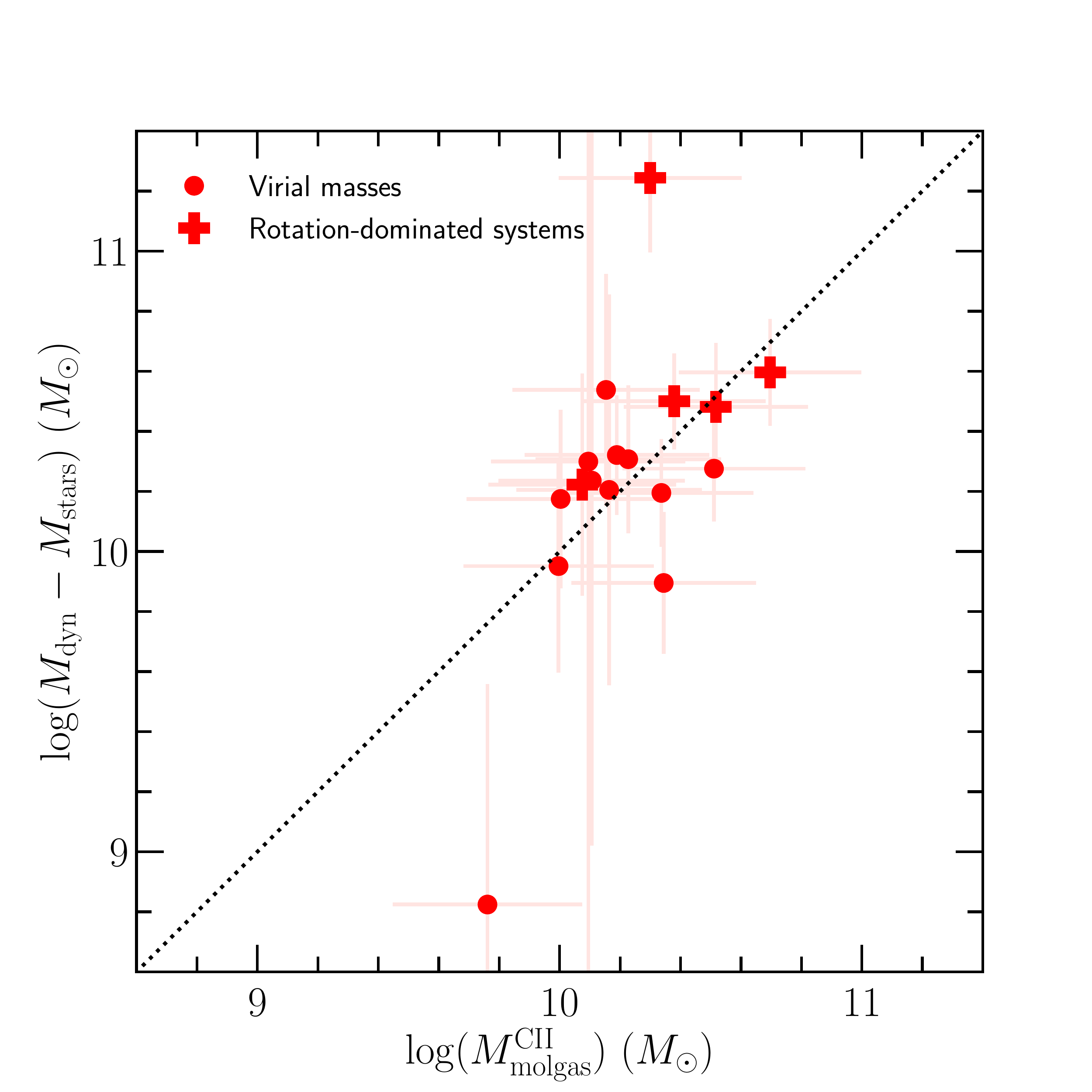}
\caption{Comparison of molecular gas masses of ALPINE non-merger galaxies as derived from the [C\,{\sc ii}] luminosity (Eq.~(\ref{eq:zanella})) and the dynamical mass after subtracting $M_{\rm stars}$ (the relative contribution of dark matter is assumed to be negligible). The dynamical masses, accessible only for the ALPINE galaxies with available [C\,{\sc ii}] size measurements \citep{fujimoto19}, are computed using the virial mass definition (Eq.~(\ref{eq:virial}); filled circles), except for 5 objects classified as rotation-dominated \citep{lefevre19} for which we consider the disk-like gas potential distribution (Eq.~(\ref{eq:disk}); crosses). The dotted line is the one-to-one relation. There is a good agreement between $M^{\rm CII}_{\rm molgas}$ and molecular gas masses inferred from dynamical masses.}
\label{fig:Mmolgas-Dynamical}
\end{figure}
%
%-----------------------------------------------------------------------

\citet{bethermin19} constructed the mean stacked FIR SEDs specific to ALPINE galaxy analogues, following the same prescriptions as in \citet{bethermin15}, but using the more recent COSMOS catalogue of \citet{davidzon17} and deep SCUBA2 data at 850~$\mu$m from \citet{casey13}. Moreover, the targets to be stacked were selected with properties analogous to the ones of the ALPINE galaxies: ${\rm SFR} \gtrsim 10~M_{\odot}~\rm yr^{-1}$, and redshift bins of $4<z<5$ and $5<z<6$. The resulting SEDs are best-fit by the \citet{bethermin17} SED template, but both the \citet{schreiber18} SED template and a modified blackbody (MBB) with dust opacity spectral index fixed to $\beta = 1.8$ and luminosity-weighted dust temperature of $41\pm 1$~K at $z<5$ and $43\pm 5$~K at $z>5$ provide a good fit \citep[$\chi^2 < 4$ for 4 degrees of freedom; see Fig.~9 in][]{bethermin19}.

We adopt the \citet{bethermin17} FIR SED template to estimate $L_{850\mu \rm m}$ of the 11 ALPINE non-merger galaxies with FIR continuum detections by scaling the measured monochromatic rest-frame 158~$\mu$m luminosity 
%to the luminosity in the long-wavelength Rayleigh-Jeans tail 
by the ratio between the SED-predicted luminosities at 850~$\mu$m and 158~$\mu$m. Using then the calibration of \citet{kaasinen19}\footnote{The calibration of \citet{kaasinen19} is comparable to the calibration of \citet{scoville16} with a constant $\alpha_{850\mu{\rm m}} = L_{\rm 850\mu \rm m}/M^{850\mu \rm m}_{\rm molgas} = (6.7\pm 1.7)\times 10^{19}~{\rm erg~s^{-1}~Hz^{-1}}~M_{\odot}^{-1}$, although it shows some deviations at $L_{\rm 850\mu \rm m}\lesssim 5\times 10^{30}~\rm erg~s^{-1}~Hz^{-1}$, but which remains well within the scatter of data used to establish the calibrations.}:
\begin{equation}
\label{eq:Scoville}
M^{850\mu \rm m}_{\rm molgas}~(M_{\odot}) = \left(\frac{L_{850\mu \rm m}}{\rm erg~s^{-1}~Hz^{-1}}\right) \left(\frac{1}{6.2\times 10^{19} (L_{\rm 850\mu \rm m}/10^{31})^{0.07}}\right),
\end{equation} 
we derive the molecular gas masses from the extrapolated rest-frame 850~$\mu$m luminosities. These values, although relying on multiple assumptions (e.g., the SED template), are independent measurements to be compared with $M_{\rm molgas}$ inferred from the [C\,{\sc ii}] luminosity. The comparison is shown in Fig.~\ref{fig:Mmolgas-Scoville}.
%These 850~$\mu$m molecular gas masses, despite their arguable reliability since fully relying on the SED template assumed for the ALPINE galaxies, represent measurements of the ISM gas masses of the ALPINE galaxies independent from the molecular gas masses inferred from the [C\,{\sc ii}] luminosity. In Fig.~\ref{fig:Mmolgas-Scoville} we compare the 850~$\mu$m and [C\,{\sc ii}] molecular gas masses (filled circles). 
Within $1-2\,\sigma$ uncertainty of $0.3-0.6$~dex, we find an agreement between these two measurements, although there is some trend for a systematic overestimate of $M^{\rm CII}_{\rm molgas}$ with respect to $M^{850\mu \rm m}_{\rm molgas}$ by 0.3~dex, on average. A similar offset is observed for the \citet{schreiber18} SED template and the MBB. On the other hand, when considering the calibration of \citet[][Table~5 and $\log (M_{\rm stars}/M_{\odot})>9$]{groves15}, obtained for local galaxies, between monochromatic luminosity in the {\it Herschel} PACS 160~$\mu$m band and gas mass, which relies on much fewer assumptions, we find only a marginal overestimate by 0.1~dex of $M^{\rm CII}_{\rm molgas}$ relative to these gas mass estimates (open squares).

The observed $M^{\rm CII}_{\rm molgas}$ overestimate with respect to $M^{850\mu \rm m}_{\rm molgas}$ may be attributed to three possible effects. First, it points to potential contributions from the neutral atomic and ionised phases to the measured [C\,{\sc ii}] emission, in addition to the molecular gas phase. Second, it suggests that the calibration of \citet{kaasinen19} may not be valid for the ALPINE galaxies at $z\gtrsim 4.5$. Remember that the \citet{scoville14} method assumes a constant dust-to-gas mass ratio of $\sim 1:100$, as also supported by \citet{kaasinen19}. Yet we may expect a lower dust-to-gas mass ratio ($\propto \alpha_{850\mu \rm m}$ in Eq.~(\ref{eq:Scoville})) for the ALPINE galaxies, 
%which would imply higher $M^{850\mu \rm m}_{\rm molgas}$, 
since ALPINE galaxies have lower $M_{\rm stars}$ (with a median of $10^{9.7}~M_{\odot}$) than SFGs with $M_{\rm stars} > 10^{10.3}~M_{\odot}$ studied by \citet{scoville16} and \citet{kaasinen19} and are deficient in dust obscured star-formation activity with respect to lower redshift SFGs as found by \citet{fudamoto20}.
%a higher conversion factor between rest-frame 850~$\mu$m luminosity and molecular gas mass for the ALPINE galaxies (Eq.~\ref{eq:Scoville}). 
To reconcile $M^{850\mu \rm m}_{\rm molgas}$ with $M^{\rm CII}_{\rm molgas}$, $\alpha_{850\mu \rm m}$ would need to be lower by a factor of $\sim 2$. Third, it supports that the SED in the Rayleigh-Jeans tail out to 850~$\mu$m rest-frame could be dominated by a cold component. When fixing the dust opacity spectral index to $\beta = 1.8$ and considering this time a mass-weighted dust temperature of $T_{\rm dust} = 25$~K in the MBB SED parametrization, similarly to \citet{scoville16,scoville17}, which we use to extrapolate the 158~$\mu$m luminosity to the 850~$\mu$m luminosity, we obtain comparable gas masses (filled stars in Fig.~\ref{fig:Mmolgas-Scoville}) to $M^{\rm CII}_{\rm molgas}$.

\subsubsection{The dynamical masses}
\label{sect:Mmolgas-Mdyn}

As discussed in \citet{lefevre19}, 2/3 of the ALPINE [C\,{\sc ii}]-detected galaxies are moderately spatially resolved. For a subset of 18 non-merger galaxies with high-SNR ($\gtrsim 5$) [C\,{\sc ii}] detections,
%non-mergers and galaxies with C$^+$ emission detected at $>5\,\sigma$, 
\citet{fujimoto19} derived their [C\,{\sc ii}] sizes by performing exponential-disk profile fits in the visibility plane with \texttt{UVMULTIFIT} \citep{martividal14}. The circularized effective radii ($r_e$), defined as the square-root of the product of the effective major and minor axes, are adopted as size measurements and are listed in Table~1 of \citet{fujimoto19}. For the ALPINE galaxies with size measurement, we can derive their dynamical mass 
%from the width of the [C\,{\sc ii}] line 
under the assumption that the gas potential structure of ALPINE galaxies arises in a virialised spherical system of radius equal to the measured circularized effective radius and with the one-dimensional velocity dispersion ($\sigma_{\rm CII}$) inferred from the full-width half maximum ($\rm FWHM_{\rm CII}^{\rm intrinsic}$) of the [C\,{\sc ii}] line corrected for final channel spacing\footnote{$\rm FWHM_{\rm CII}^{\rm intrinsic} = \sqrt{\rm FWHM_{\rm CII}^{\rm observed~2}-25^2}$, where $25~\rm km~s^{-1}$ is the spectral resolution that, in our spectral configuration, equals the final channel spacing (ALMA Technical Handbook). We then obtain the velocity dispersion from $\sigma_{\rm CII} = \rm FWHM_{\rm CII}^{\rm intrinsic}/\sqrt{8\ln(2)}$.}:
\begin{equation}
\label{eq:virial}
M_{\rm dyn}^{\rm virial}~(M_{\odot}) = 1.56\times 10^6 \left(\frac{\sigma_{\rm CII}}{{\rm km~s^{-1}}}\right)^2 \left(\frac{r_e}{{\rm kpc}}\right),
\end{equation}
following Eq.~(10) in \citet{bothwell13}. This virialised spherical geometry dynamical mass is 0.13~dex larger than the dynamical mass we would obtain if we assumed a disk-like gas potential distribution for the same source size, the same $\rm FWHM_{\rm CII}^{\rm intrinsic}$, and a mean inclination angle of the source of $\langle \sin i \rangle = \pi/4$ \citep{law09,wang13,capak15}. But, the virial mass has the advantage of avoiding to add the supplementary uncertainty on the source orientation needed in the computation of the dynamical mass for disk geometry. For 5 out of the 9 ALPINE galaxies classified as rotation-dominated systems \citep{lefevre19}, we obtained robust [C\,{\sc ii}] minor and major axis ratio measurements \citep{fujimoto19}, which enable us to constrain their disk inclination angle ($i$) as $i = \cos^{-1}(\rm minor/major)$. For these 5 galaxies, we also compute their dynamical mass for the disk-like gas potential distribution:
\begin{equation}
\label{eq:disk}
M_{\rm dyn}^{\rm rotation}~(M_{\odot}) = 1.16\times 10^5 \left(\frac{\upsilon_{\rm cir}}{{\rm km~s^{-1}}}\right) \left(\frac{2 r_e}{{\rm kpc}}\right),
\end{equation}
where the circular velocity of the gaseous disk is $\upsilon_{\rm cir} = 1.763 \sigma_{\rm CII}\,/\sin(i)$. The corresponding dynamical masses are randomly scattered by up to $\pm 0.25$~dex from virial masses.

Since the relative contribution of dark matter in the internal regions of galaxies (at $<(1-2) r_e$) is expected to be low (\citet{barnabe12} report a dark matter fraction within $2.2r_e$ of at most $0.28^{+0.15}_{-0.10}$), the dynamical mass may be assumed to reflect the total baryonic mass, which can be used to obtain an estimate of $M_{\rm molgas}$ after subtracting $M_{\rm stars}$. Out of the 18 ALPINE non-merger galaxies with size measurements, for one galaxy\footnote{In VUDS~COSMOS~5101218326 the virial mass is smaller than $M_{\rm stars}$, likely because of an overestimation of $M_{\rm stars}$ given the distorted morphology of the galaxy in the {\it Hubble} Space Telescope optical bands \citep{faisst19}, although we cannot exclude an underestimation of its virial mass as well.} the virial mass ends up to be smaller than its $M_{\rm stars}$. For the remaining 17 ALPINE galaxies, we can cross-match the molecular gas masses obtained from their dynamical and stellar masses with the gas masses inferred from their [C\,{\sc ii}] luminosity. For 12 ALPINE galaxies we consider the virial masses, and for the 5 ALPINE galaxies classified as rotation-dominated we consider the dynamical masses derived for a disk-like gas potential. As shown in Fig.~\ref{fig:Mmolgas-Dynamical}, there is a good agreement within the $1\,\sigma$ uncertainty of 0.3~dex between these respective molecular gas mass estimates, except for two outliers (they do not show any systematic trend). 
%At the end, we might expect 
A good one-to-one relationship between the two molecular gas mass estimates was expected, since the dynamical mass, if tracing the baryonic mass only (no dark matter), accounts for $M_{\rm stars}$ plus the total gas mass that includes all gas phases (molecular, atomic, ionised) as likely does the [C\,{\sc ii}] emission.

%
%-----------------------------------------------------------------------

\section{Comparison sample}
\label{sect:compilation}

Tremendous observational efforts have been dedicated to determine the molecular gas content of galaxies from the present day to high redshift, using either the CO emission or the FIR dust continuum as molecular gas mass tracers. These tracers have their respective strengths and uncertainties \citep[see][]{bolatto13,genzel15,scoville16,tacconi18}. While the former tracer is the most commonly used and well-calibrated tracer in the local Universe, the latter tracer, which usually relies on a single-band measurement preferably in the Rayleigh-Jeans tail of the FIR SED, is particularly cheap in terms of ALMA observing time. Here we propose to compare the ALPINE $M^{\rm CII}_{\rm molgas}$ with a compilation of local to high-redshift MS SFGs with molecular gas masses derived from CO luminosity measurements reported in the literature. 

We build up the database of CO-detected MS SFGs starting from the exhaustive compilation of CO luminosity measurements in MS SFGs at $z>1$ presented in \citet{dessauges15,dessauges17}. We extend this compilation with CO measurements at $z>1$ published since 2015\footnote{We do not include the CO detection of \citet{dodorico18} in our compilation of CO-detected MS SFGs, because $M_{\rm stars}$ of the corresponding galaxy is not known.} by \citet{seko16}, \citet{papovich16}, \citet{gonzalez17}, \citet{magdis17}, \citet{valentino18}, \citet{gowardhan19}, \citet{kaasinen19}, \citet{molina19}, \citet{aravena19}, \citet{bourne19}, \citet{pavesi19}, and \citet{cassata20}. And we include the release of the NOEMA PHIBSS2 legacy survey at $0.5<z<2.5$, described in \citet{tacconi18} and \citet{freundlich19}. We adopt the MS parametrization from \citet[][Eq.~(28)]{speagle14}, similarly to what was done for PHIBSS2, and retain only SFGs lying within the MS dispersion of $\Delta {\rm MS} = \log({\rm SFR/SFR_{MS}}) = \pm 0.3$~dex. The updated compilation comprises a total of 101 CO luminosity measurements of MS SFGs at $1<z<3.7$ 
%(190 at $0.1 < z < 3.7$) 
and with $M_{\rm stars} = 10^{9.5}-10^{11.7}~M_{\odot}$, plus the CO detection at $z=5.65$ from \citet{pavesi19}, but is still under-sampled at high-redshift ($z>2.5$) and at the low-$M_{\rm stars}$ end ($M_{\rm stars} < 10^{10}~M_{\odot}$). The compilation of \citet{dessauges15} also contained CO(1--0) measurements for a non-exhaustive number of local spiral galaxies and MS SFGs at $z<1$. We now add the CO(1--0) measurements from the final xCOLD GASS survey at $0.01<z<0.05$ performed with the IRAM 30~m telescope \citep{saintonge16,saintonge17}, which now extends to lower $M_{\rm stars}$ than in previous samples, i.e.\ $\log(M_{\rm stars}/M_{\odot}) = 9-10$.

At $z>0.5$, the CO(1--0) transition is often replaced by a high-$J$ CO transition with $J=2$ to 5, which requires the calibration of temperature and density from the CO spectral line energy distribution (CO SLED) to access the CO luminosity correction factor $r_{J,1} = L'_{{\rm CO}(J\rightarrow J-1)}/L'_{\rm CO(1-0)}$. CO SLED observations in MS SFGs at $z\sim 1-3.7$ converge toward $r_{2,1} = 0.81\pm 0.15$, $r_{3,1} = 0.57\pm 0.11$, $r_{4,1} = 0.33\pm 0.06$, and $r_{5,1} = 0.23\pm 0.04$ \citep{daddi15,dessauges15,dessauges19,cassata20}. In order to have a homogeneous comparison sample, we adopt these CO luminosity correction factors to all CO $J$\,$\rightarrow$\,$J-1$ luminosity measurements in our compilation, and we derive the molecular gas masses, $M_{\rm molgas} = \alpha_{\rm CO}^Z \left(\frac{L'_{{\rm CO}(J\rightarrow J-1)}}{r_{J,1}}\right)$, assuming the same CO-to-H$_2$ metallicity-dependent conversion function:
\begin{align}
\alpha_{\rm CO}^Z & ~(M_{\odot} ({\rm K~km~s^{-1}~pc^2})^{-1}) = \alpha_{\rm CO,MW}~\times \notag \\
& \sqrt{0.67 \exp\left(0.36\times 10^{-(12+\log({\rm O/H})-8.67)}\right)}~\times \notag \\ 
& \sqrt{10^{-1.27 (12+\log({\rm O/H})-8.67)}},
\end{align}
which corresponds to the geometrical mean of the metallicity-dependent conversion functions of \citet{bolatto13} and \citet{genzel12}, following Eq.~(2) in \citet{tacconi18}. We adopt the Milky Way CO-to-H$_2$ conversion factor of \citet{strong96}, $\alpha_{\rm CO,MW} = 4.36~M_{\odot} ({\rm K~km~s^{-1}~pc^2})^{-1}$, which includes the correction factor of 1.36 for helium. To estimate the metallicities of the CO-detected SFGs when direct metallicity measurements are not available, we use the redshift-dependent mass-metallicity relation defined by \citet{genzel15}\footnote{The redshift-dependent mass-metallicity relation of \citet{genzel15} was constructed by combining mass–metallicity relations at different redshifts presented by \citet{erb06}, \citet{maiolino08}, \citet{zahid14}, and \citet{wuyts14}.}, calibrated to the \citet{pettini04} metallicity scale and the solar abundance of $12+\log({\rm O/H})_{\odot} = 8.67$ \citep{asplund04}:
\begin{equation}
\label{eq:mass-metallicity}
12+\log({\rm O/H})_{\rm PP04} = a-0.087 (\log(M_{\rm stars})-b)^2
\end{equation}
with $a = 8.74$ and $b = 10.4+4.46\log(1+z)-1.78(\log(1+z))^2$. As discussed in \citet{dessauges17}, $\alpha_{\rm CO}^Z$ increases with redshift for any given $M_{\rm stars}$, and at any given redshift increases with decreasing $M_{\rm stars}$. As a result, $\alpha_{\rm CO}^Z$ might be particularly uncertain at high redshifts ($z\gtrsim 3$) and for small $M_{\rm stars}$ ($M_{\rm stars} \lesssim 10^{10}~M_{\odot}$), because of the less constrained mass-metallicity relation in this range of physical parameters.

%
%-----------------------------------------------------------------------
\begin{figure}
\centering
\includegraphics[width=8.5cm,clip]{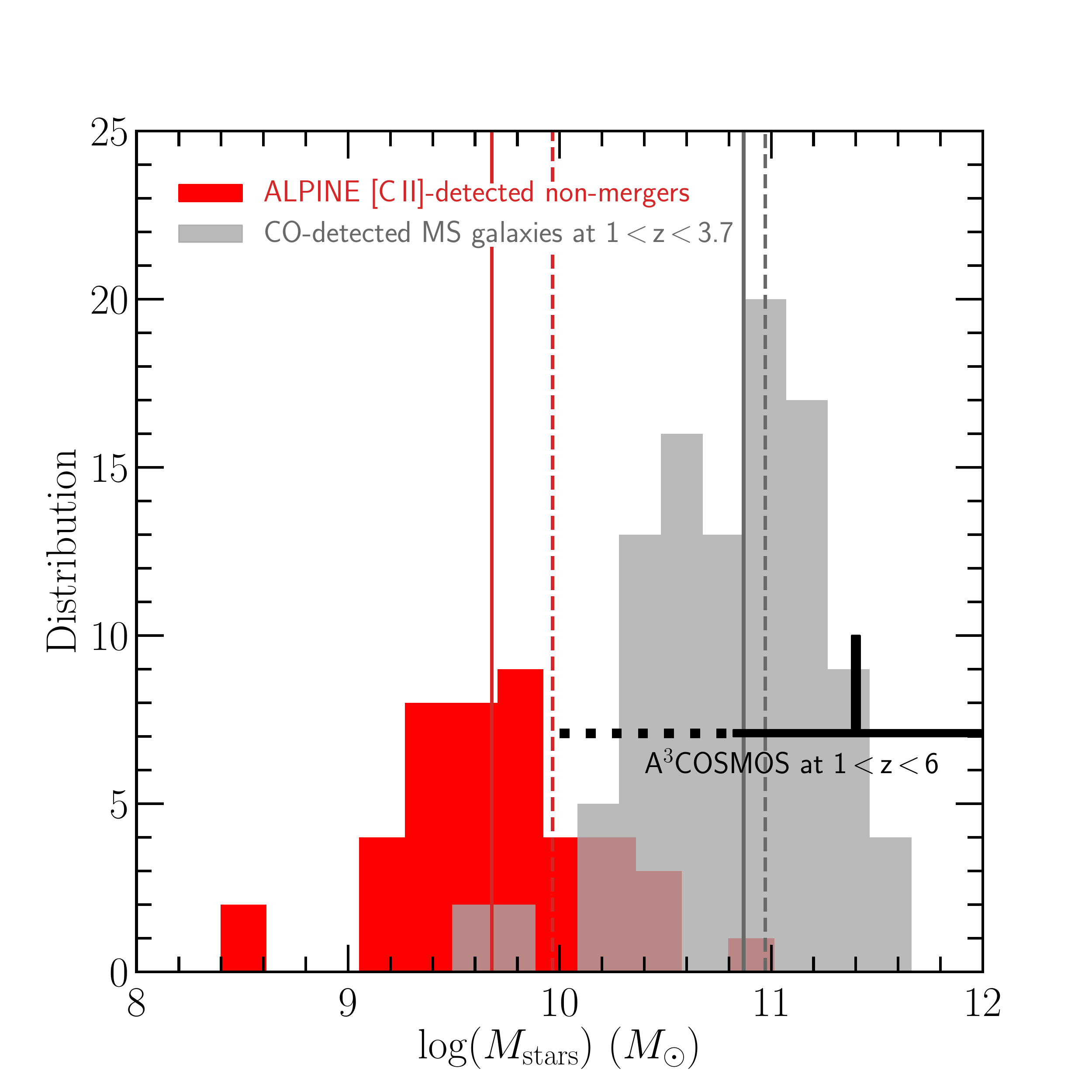}
\caption{Distribution of stellar masses of the 44 ALPINE [C\,{\sc ii}]-detected non-merger galaxies at $z=4.4-5.9$ (red histogram) and the comparison sample of 101 CO-detected MS SFGs at $1<z<3.7$ compiled from the literature (grey histogram). The solid and dashed lines correspond, respectively, to the medians and means of the two distributions. The black thick lines show the range and the mean of $M_{\rm stars}$ of the A$^3$COSMOS galaxies at $1<z<6$ \citep{liu19b}. Clearly, the ALPINE sample probes a much lower $M_{\rm stars}$ range than previous galaxy samples with molecular gas mass measurements obtained mostly at lower redshifts.}
\label{fig:Mstars-distributions}
\end{figure}
%
%-----------------------------------------------------------------------

Finally, to check if our compilation of high-redshift SFGs at $0.1<z<3.7$ with molecular gas masses derived from CO luminosity measurements is representative of MS SFGs at these redshifts, we consider the mean $M_{\rm molgas}$ obtained by \citet{bethermin15} from their stacking analysis of the IR to millimeter emission of MS SFGs, with an average $M_{\rm stars}$ of $\sim 10^{10.8}~M_{\odot}$, blindly selected in the COSMOS field between $z=0.25$ and $z=4$. For a coherent comparison, we rescale the molecular gas masses of \citet{bethermin15} to the mass-metallicity relation used in the CO compilation (Eq.~(\ref{eq:mass-metallicity})). Nevertheless, we keep the metallicity correction of $0.3\times (1.7-z)$~dex they applied at $z>1.7$ and which becomes significant for galaxies beyond $z>2.5$. We find that the respective molecular gas depletion timescales and gas fractions globally agree, supporting that the sample of CO-detected SFGs is unbiased, except maybe in the redshift bin of $1<z<1.5$ where the CO-measured molecular gas masses tend to be higher than the \citet{bethermin15} FIR SED stack results (see Fig.~\ref{fig:tdepl}, left panel and Fig.~\ref{fig:fmolgas}, top panel).

Recently, \citet{liu19b} published $M_{\rm molgas}$ measurements for about 700 galaxies at $0.3<z<6$, extracted on an automated prior-based and blind-based ALMA Archive mining in the COSMOS field (hereafter A$^3$COSMOS, with spectroscopic redshifts available for 36\% of the sample (\citeauthor{liu19a} \citeyear{liu19a})). The molecular gas masses were derived from single-band FIR continuum and multi-wavelength FIR SEDs. The A$^3$COSMOS galaxies, however, are mostly probing the high $M_{\rm stars}$ domain of MS SFGs at $z>1$ with $M_{\rm stars}\sim 10^{11}-10^{12}~M_{\odot}$. They thus are, on average, $10-100$ times more massive than the ALPINE [C\,{\sc ii}]-detected galaxies that have a median $M_{\rm stars}$ of $\sim 10^{9.7}~M_{\odot}$ (and a mean of $\sim 10^{10}~M_{\odot}$). Consequently, in terms of the respective $M_{\rm stars}$ distributions shown in Fig.~\ref{fig:Mstars-distributions}, our compilation of CO-detected MS galaxies at $z>1$ represents a better comparison sample for the ALPINE galaxies, despite the fact that in the redshift range of ALPINE galaxies ($z=4.4-5.9$) one single CO detection is included, against 24 $M_{\rm molgas}$ measurements for MS SFGs in the A$^3$COSMOS sample. With a median $M_{\rm stars}$ of $\sim 10^{10.9}~M_{\odot}$ (and a mean of $\sim 10^{11}~M_{\odot}$), the CO-detected SFGs globally have adequate masses at $1<z<3.7$ to plausibly be the descendants of the ALPINE galaxies according to the multi-epoch abundance matching simulations \citep{behroozi13,behroozi19,moster13,moster18}, as  discussed in Sect.~\ref{sect:fmolgas-evolution}.

%
%-----------------------------------------------------------------------

\section{Analysis and discussion}
\label{sect:discussion}

The comparison of the molecular gas masses inferred from the [C\,{\sc ii}] luminosity, following the calibration proposed by \citet{zanella18}, with three independent gas mass tracers discussed in Sect.~\ref{sect:Mmolgas} yields reassuring results supporting [C\,{\sc ii}] as a statistically reliable tracer of cold molecular gas for the ALPINE galaxies. %within $1\,\sigma$ uncertainty. 
The [C\,{\sc ii}]-estimated gas masses are associated with large error bars ($\sim 0.3$~dex), but these error bars appear to be comparable to those of other tracers. In what follows, we adopt the [C\,{\sc ii}] gas masses derived for the ALPINE galaxies to study the evolution of the molecular gas content of MS SFGs up to $z\sim 6$. We stress that if instead we choose another of the tested molecular gas mass tracers, we would obtain similar conclusions.
%but we stress that using either of the tested molecular gas mass tracers will not affect our findings.

%
%-----------------------------------------------------------------------

\subsection{Molecular gas depletion timescale}
\label{sect:tdepl}

\begin{figure*}
\centering
\includegraphics[width=6.3cm,clip]{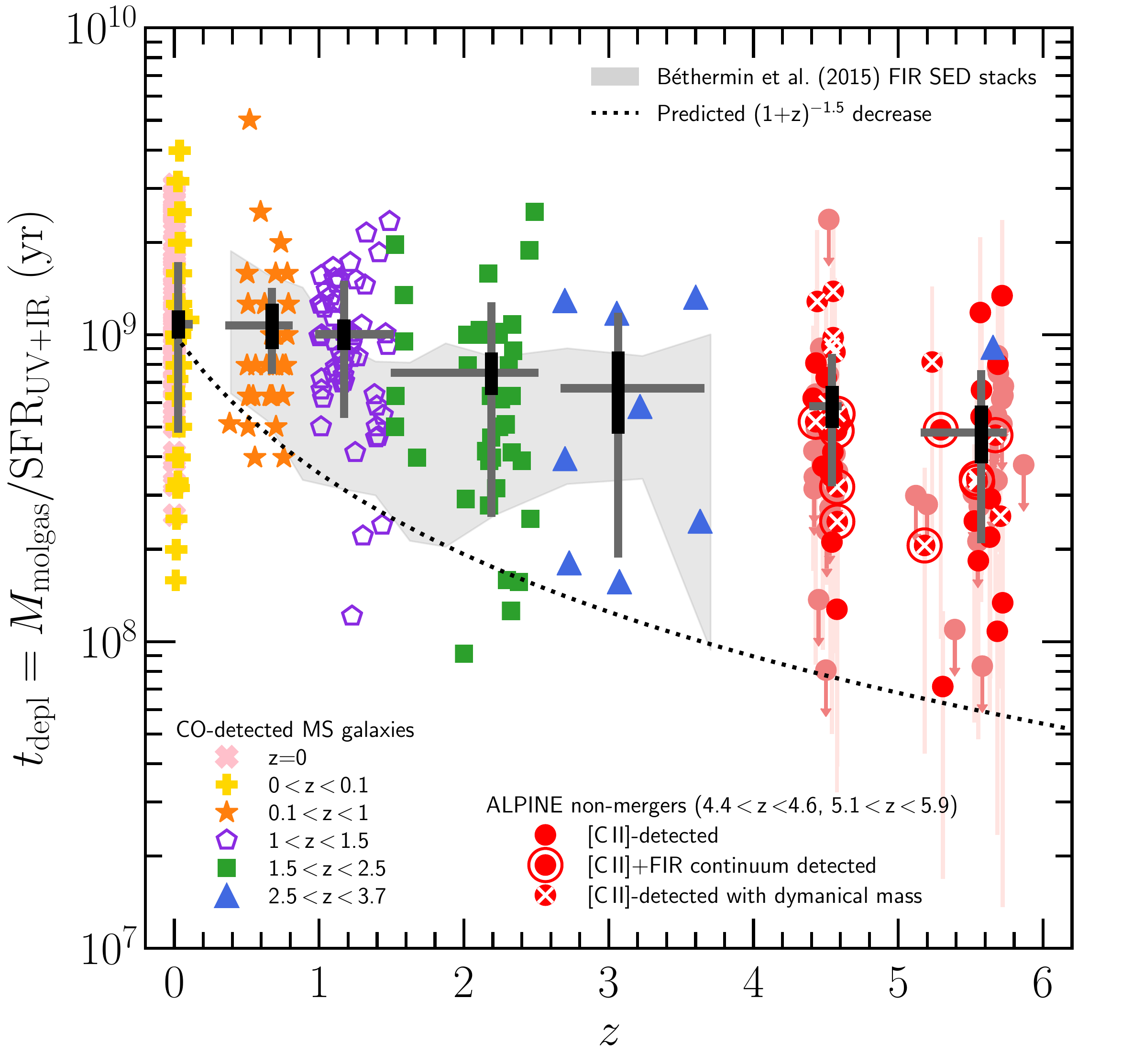}
\includegraphics[width=5.95cm,clip]{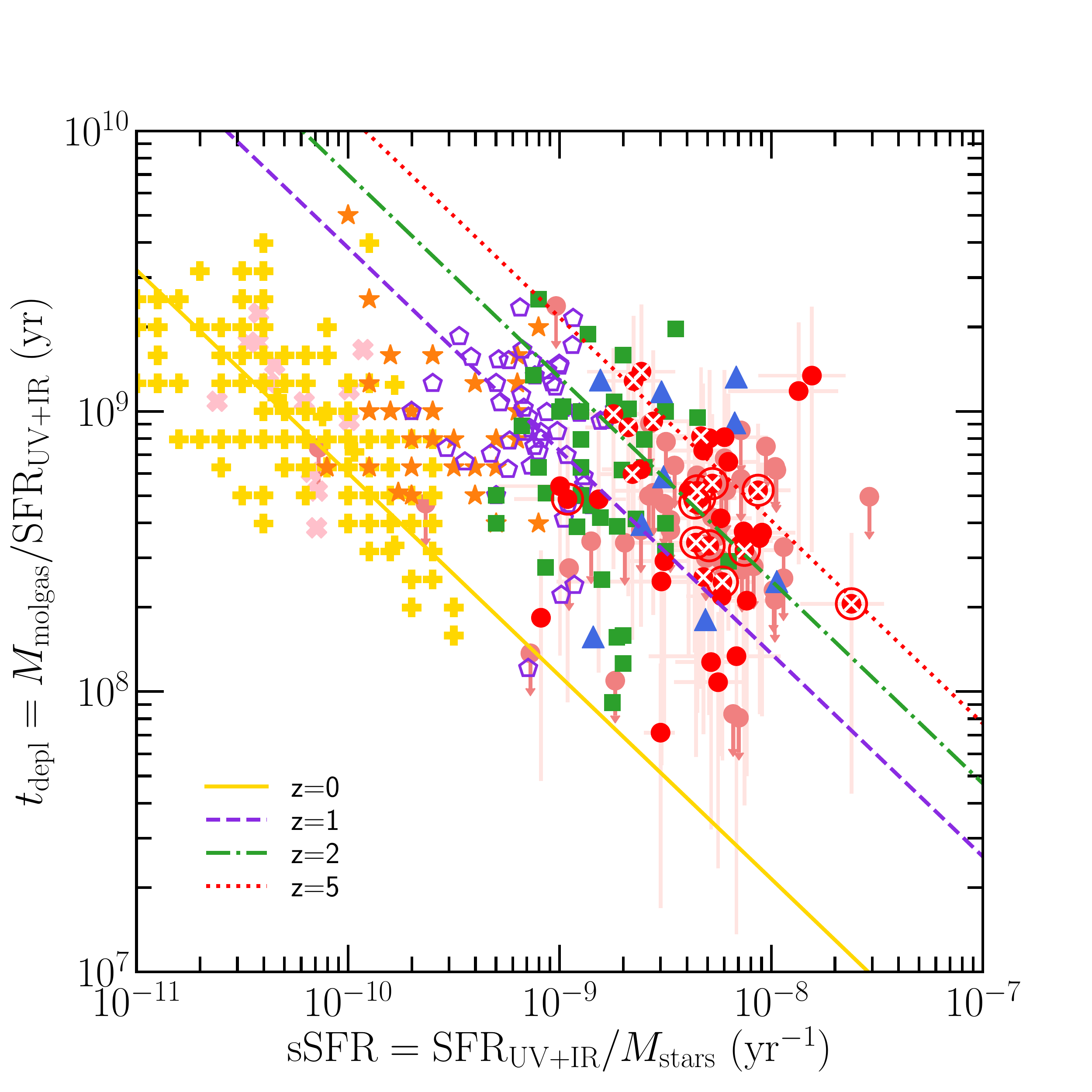}
\includegraphics[width=5.95cm,clip]{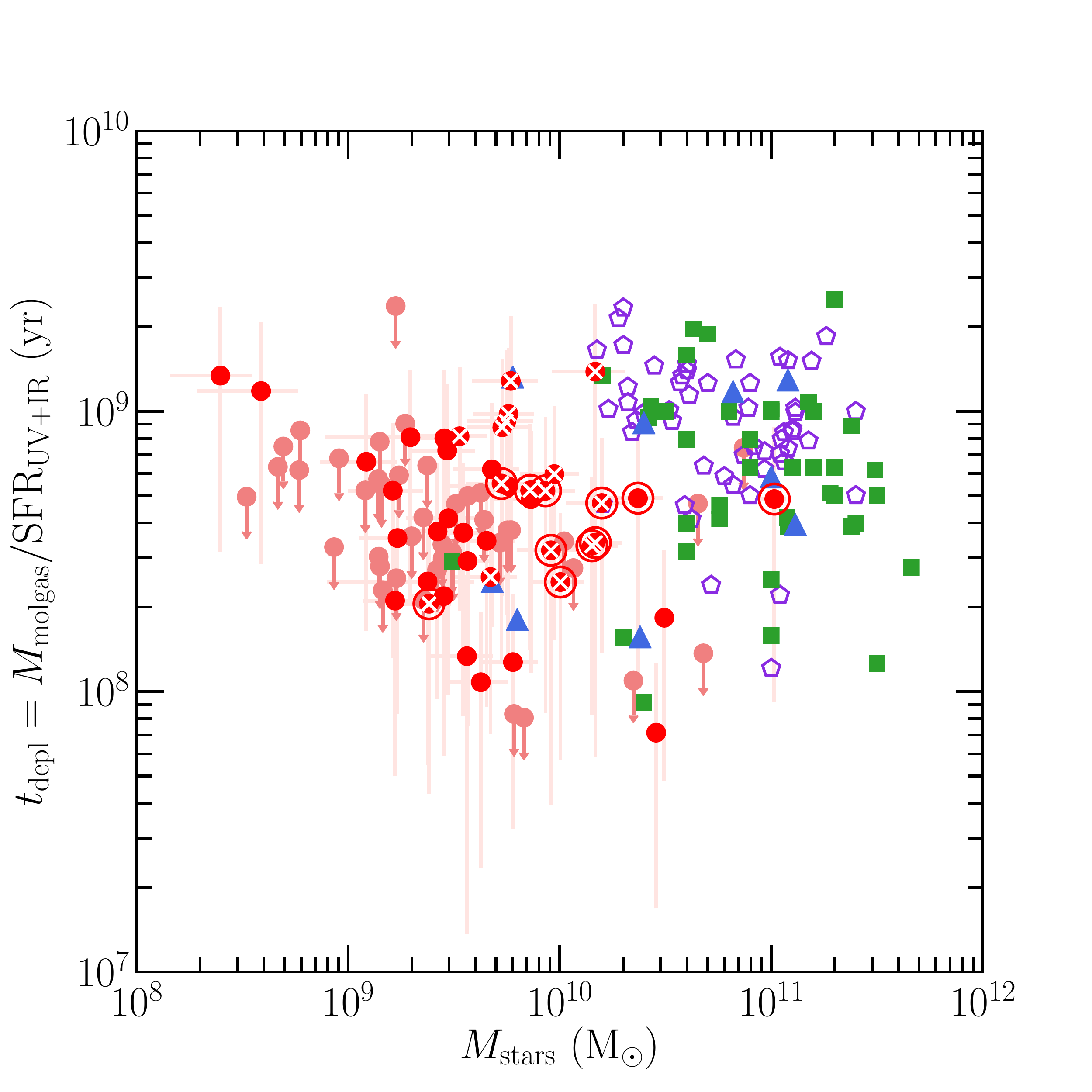}
\caption{Molecular gas depletion timescales plotted for the ALPINE non-merger galaxies distributed in two redshift bins of $4.4<z<4.6$ and $5.1<z<5.9$ (red circles; encircled red circles mark the ALPINE galaxies detected in FIR dust continuum; crossed red circles mark the ALPINE galaxies with dynamical mass measurements; and light-red arrows correspond to $3\,\sigma$ upper limits) and for our compilation of CO-detected MS SFGs from the literature color-coded in 6 redshift bins of $z=0$ (pink crosses), $0<z<0.1$ (yellow pluses), $0.1<z<1$ (orange stars), $1<z<1.5$ (violet open pentagons), $1.5<z<2.5$ (green squares), and $2.5<z<3.7$ (blue triangles, plus the \citet{pavesi19} object at $z=5.65$). 
{\it Left panel.} Molecular gas depletion timescales shown as a function of redshift. The respective means, errors on the mean, and standard deviations per redshift bin are indicated by the black/grey big crosses. The light-grey shaded area corresponds to the depletion timescales obtained by \citet{bethermin15} from FIR SED stacks. The $t_{\rm depl}$ means per redshift bin follow a decrease out to $z\sim 6$, but much shallower than the $(1+z)^{-1.5}$ decline predicted in the framework of the bathtub model (dotted line).
{\it Middle panel.} Molecular depletion timescales shown as a function of specific star formation rate. A strong anti-correlation between $t_{\rm depl}$ and sSFR is observed at $z=0$ \citep[yellow solid line from][]{saintonge11} and at high redshift. The displacement along the sSFR-axis for galaxies at higher redshifts is compatible with the sSFR evolution with redshift (violet dashed line at $z=1$, green dashed-dotted line at $z=2$, and red dotted line at $z=5$, as computed using the ${\rm sSFR}(z)$ parametrization from \citeauthor{speagle14} \citeyear{speagle14}, Eq.~(28)).
{\it Right panel.} Molecular depletion timescales, restricted to $z\sim 1-5.9$ SFGs, shown as a function of $M_{\rm stars}$. No correlation between $t_{\rm depl}$ and $M_{\rm stars}$ is observed for SFGs at $z\sim 1-5.9$.}
\label{fig:tdepl}
\end{figure*}

From estimates of $M_{\rm molgas}$ and SFR of galaxies, we can infer their molecular gas depletion timescale, defined as $t_{\rm depl} = M_{\rm molgas}/{\rm SFR}$. This gas depletion timescale (or gas consumption timescale) describes how long each galaxy may sustain star formation at the measured rate before running out of molecular gas fuel under the assumption that the gas reservoir is not replenished. Since the earliest CO luminosity measurements in high-redshift MS SFGs, there has been evidence for shorter $t_{\rm depl}$ at high redshift, such that $t_{\rm depl}\sim 1-2$~Gyr observed at $z=0$ \citep[e.g.,][]{bigiel08,leroy13,saintonge17} drops by a factor of $\sim 2$ at $z\sim 2.5$ \citep[e.g.,][]{tacconi13,tacconi18,tacconi20,saintonge13,genzel15,bethermin15,
dessauges15,dessauges17,schinnerer16,scoville17,liu19b}. Shorter $t_{\rm depl}$ correspond to higher star formation efficiencies (${\rm SFE} = 1/t_{\rm depl}$) that are taking place in high-redshift galaxies, efficient enough to exhaust similar and even larger gas reservoirs over a shorter timescale than in nearby MS SFGs. The so-far inferred $t_{\rm depl}$ evolution with redshift, up to $z\sim 3.5$, nevertheless appears much shallower than $t_{\rm depl} \sim (1+z)^{-1.5}$ (see Fig.~\ref{fig:tdepl}, left panel) that is predicted by semi-analytical and cosmological simulations developed in the framework of the bathtub model \citep[e.g.][]{dave11,dave12,genel14,lagos15}. This suggests that distant galaxies either intrinsically do not have such high SFE, or are more gas-rich than predicted, or outflows, if highly mass loaded, contribute to reduce the gas.

The ALPINE sample enables us, for the first time, to explore the $t_{\rm depl}$ evolution beyond $z\gtrsim 4.5$ for a statistically significant number of MS SFGs with a median $M_{\rm stars}$ of $10^{9.7}~M_{\odot}$. The measured $t_{\rm depl}$ means and errors in two redshifts bins of $4.4<z<4.6$ and $5.1<z<5.9$ are listed in Table~\ref{tab:tdeplfmolgas}. We provide both the means obtained when considering only 
%the [C\,{\sc ii}] detections of the 44 non-merger galaxies 
the 44 [C\,{\sc ii}]-detected galaxies and when also taking into account the ``secure'' $3\,\sigma$ upper limits of the 43 galaxies undetected in [C\,{\sc ii}] (see Sect.~\ref{sect:observations}). The latter means are computed using the survival analysis \citep[with routines described in][]{isobe86}. In particular, we use the Kaplan-Meier estimator, an unbiased non-parametric maximum likelihood estimator that determines the characteristic of a parent population with no assumption on the distribution of the parent population from which the censored sample is drawn \cite[see also][]{talia20}. The respective $t_{\rm depl}$ means without/with limits differ by about a factor of 2. 

%
%-----------------------------------------------------------------------
\begin{table}
\caption{ALPINE molecular gas depletion timescale and molecular gas fraction means in two redshift bins}             
\label{tab:tdeplfmolgas} 
\centering 
\begin{tabular}{l c c} 
\hline\hline

 & $4.4<z<4.6$ & $5.1<z<5.9$ \\
\hline  
$\langle t_{\rm depl}\rangle$ detections          & $5.8\pm 0.6$ & $4.6\pm 0.8$ \\
$\langle t_{\rm depl}\rangle$ detections+limits   & $2.3\pm 0.3$ & $2.3\pm 0.4$ \\
$\langle f_{\rm molgas}\rangle$ detections        & $0.67\pm 0.03$ & $0.59\pm 0.05$ \\
$\langle f_{\rm molgas}\rangle$ detections+limits & $0.46\pm 0.05$ & $0.46\pm 0.05$ \\
\hline     
\end{tabular}
\tablefoot{$\langle t_{\rm depl}\rangle$ values are in $10^8$~yr. The detections refer to the 44 ALPINE [C\,{\sc ii}]-detected non-merger galaxies and the limits to the ``secure'' $3\,\sigma$ upper limits of the 43 [C\,{\sc ii}]-non-detected galaxies (see Sect.~\ref{sect:observations}).}
\end{table}
%
%-----------------------------------------------------------------------
%
%-----------------------------------------------------------------------

In Fig.~\ref{fig:tdepl} (left panel) we show the molecular gas depletion timescale as a function of redshift for the ALPINE [C\,{\sc ii}]-detected non-merger galaxies (red circles) and [C\,{\sc ii}]-non-detected galaxies (light-red arrows) distributed in the redshift bins of $4.4<z<4.6$ and $5.1<z<5.9$, and for our compilation of CO-detected MS SFGs from the literature separated in 6 redshift bins of $z=0$, $0<z<0.1$, $0.1<z<1$, $1<z<1.5$, $1.5<z<2.5$, and $2.5<z<3.7$, chosen in the way that the three bins between $z=0.1$ and $z=2.5$ contain a comparable number of galaxies ($\sim 40$).
%and our compilation of CO-detected main-sequence SFGs from the literature separated in 6 redshift bins of $z=0$ (black thin crosses), $0<z<0.1$ (cyan thick crosses), $0.1<z<1$ (yellow stars), $1<z<1.5$ (blue open pentagons), $1.5<z<2.5$ (grey squares), and $2.5<z<3.7$ plus $z=5.65$\footnote{The redshift of the \citet{pavesi19} CO-detected galaxy.} (green triangles), with their respective means, errors on the mean, and standard deviations (black/grey big crosses).
We then compute the respective means, errors on the mean, and standard deviations per redshift bin (black/grey big crosses). We show the ALPINE means obtained for the 44 [C\,{\sc ii}] detections (see Table~\ref{tab:tdeplfmolgas}). We also overplot the depletion timescales obtained by \citet{bethermin15} from FIR SED stacks (light-grey shaded area). 
%These are in agreement with the $t_{\rm depl}$ means derived from CO luminosity measurements, which supports that the CO detections are not biased toward galaxies with high molecular gas masses. 
We observe a continuous decline of $t_{\rm depl}$ from $z=0$ to $z=5.9$.
%with no break at $z\gtrsim 4.5$. 
The decline follows a power-law with a slope clearly shallower than $(1+z)^{-1.5}$ (dotted line), as this latter would imply $t_{\rm depl} = 6.0\times 10^7$~yr at $z=5.5$ when fixing the zero-point at $z=0$ to 1~Gyr. This predicted $t_{\rm depl}$ value is comparable to the smallest ALPINE $t_{\rm depl}$ measurement, but is almost one order of magnitude shorter than the mean $t_{\rm depl}$ of $(4.6\pm 0.8)\times 10^8$~yr of the ALPINE [C\,{\sc ii}]-detected non-merger galaxies in the redshift bin of $5.1<z<5.9$. Even if, for the ALPINE galaxies undetected in the FIR continuum emission, we add to their measured $\rm SFR_{UV}$ the possible $\rm SFR_{IR}$ contribution, estimated using the ALPINE IRX-$\beta$ relation obtained from stacking \citep{fudamoto20} as discussed in Sect.~\ref{sect:observations}, the resulting mean $t_{\rm depl}$ of $\sim 3.8\times 10^8$~yr over $4.4<z<5.9$ is still too long compared to the $(1+z)^{-1.5}$ decline. When taking into account the ``secure'' $3\,\sigma$ upper limits of the ALPINE galaxies undetected in [C\,{\sc ii}], the mean $t_{\rm depl}$ drops to $(2.3\pm 0.4)\times 10^8$~yr in the redshift bin of 5.1 < z < 5.9. This drop suggests a steeper $t_{\rm depl}$ decrease with redshift than shown by [C\,{\sc ii}] detections, but the reached mean $t_{\rm depl}$ value is still a factor of $\sim 4$ longer than for the predicted one. Consequently, on average, MS SFGs at $z\gtrsim 4.5$ are not considerably more efficient in forming stars than those galaxies at $z\sim 2-3$,
%not that efficient in forming stars 
as also supported by the low SFE obtained by \citet{pavesi19} from the CO(2--1) molecular gas mass measurement in a MS SFG at $z=5.65$ (see the blue triangle at $z=5.65$ in the left panel of Fig.~\ref{fig:tdepl}).
%$3.7\times 10^8$~yr for the ALPINE galaxies at $4.4<z<4.6$
%$4.2\times 10^8$~yr for the ALPINE galaxies at $5.1<z<5.9$

%
%-----------------------------------------------------------------------
\begin{figure*}
\centering
\includegraphics[width=8cm,clip]{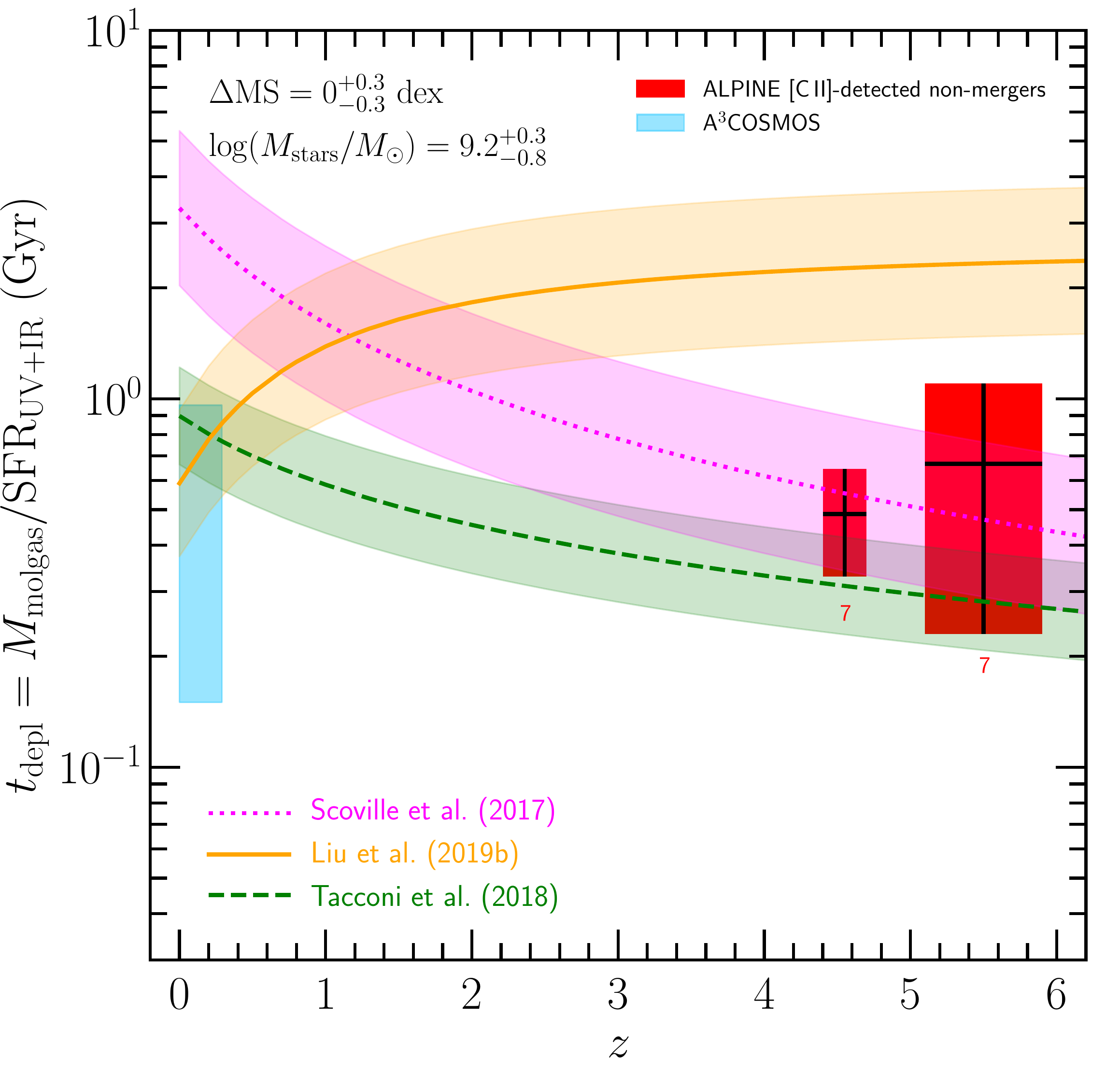} \hspace{0.2cm}
\includegraphics[width=8cm,clip]{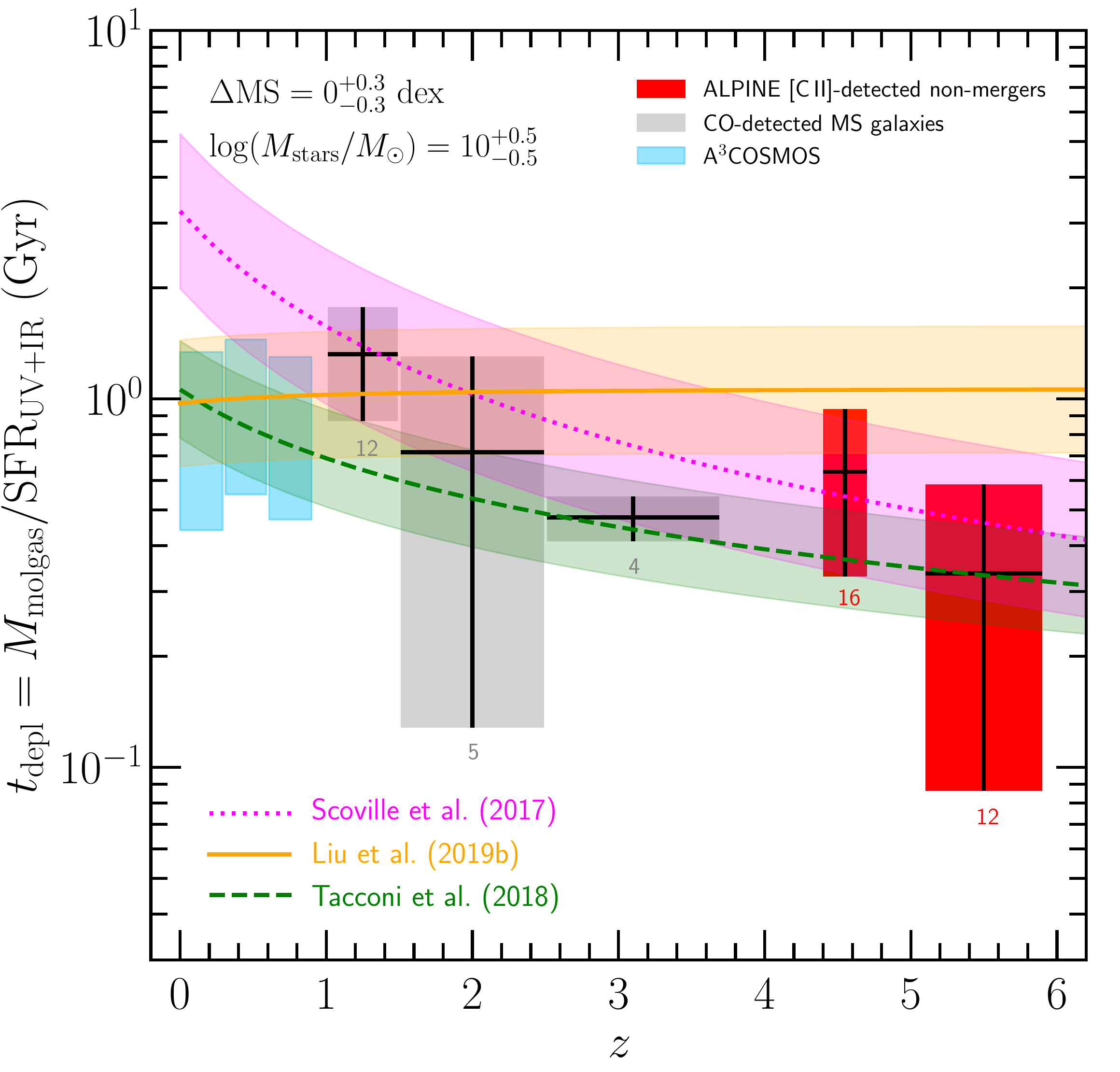}
\includegraphics[width=8cm,clip]{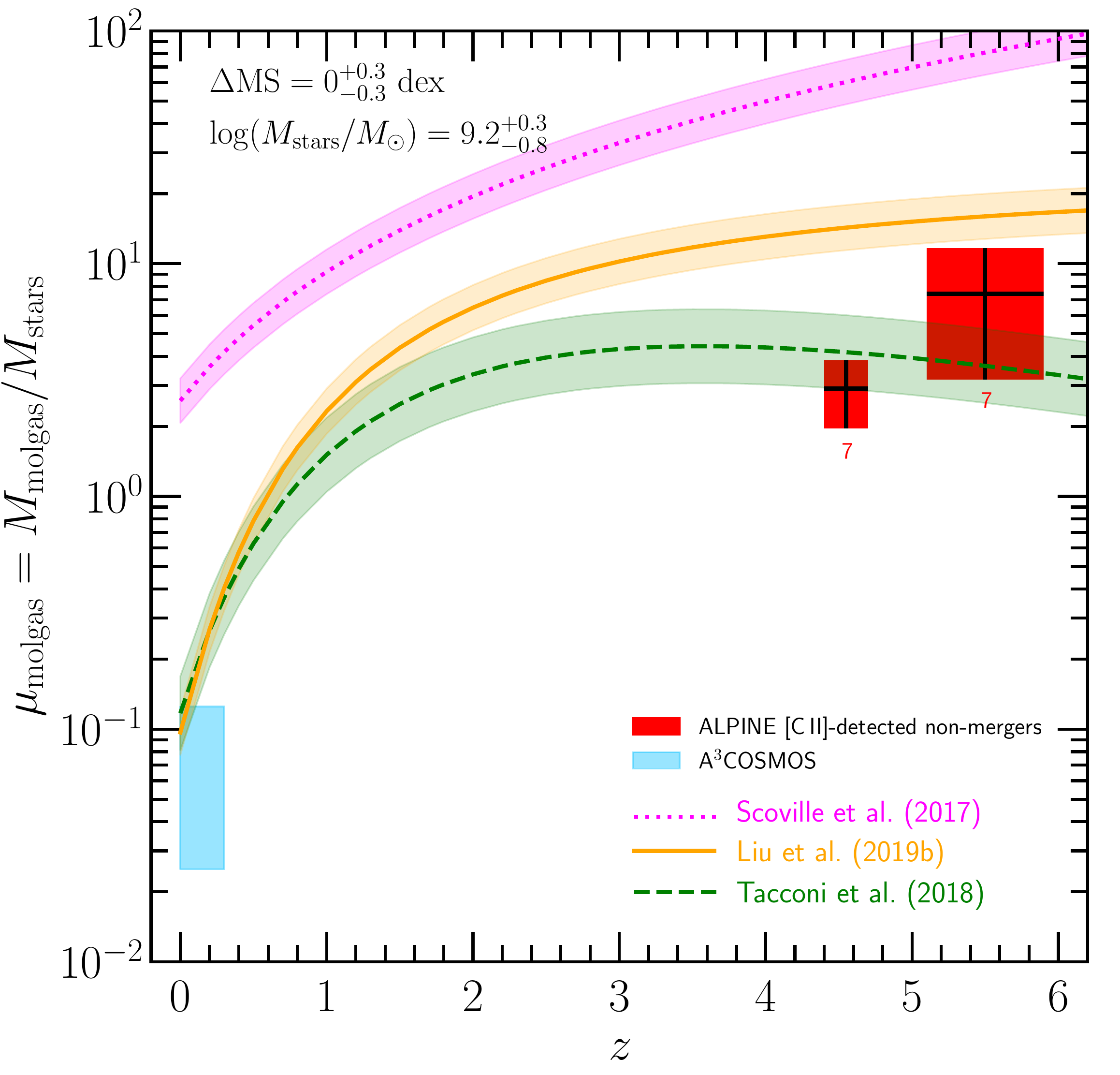} \hspace{0.2cm}
\includegraphics[width=8cm,clip]{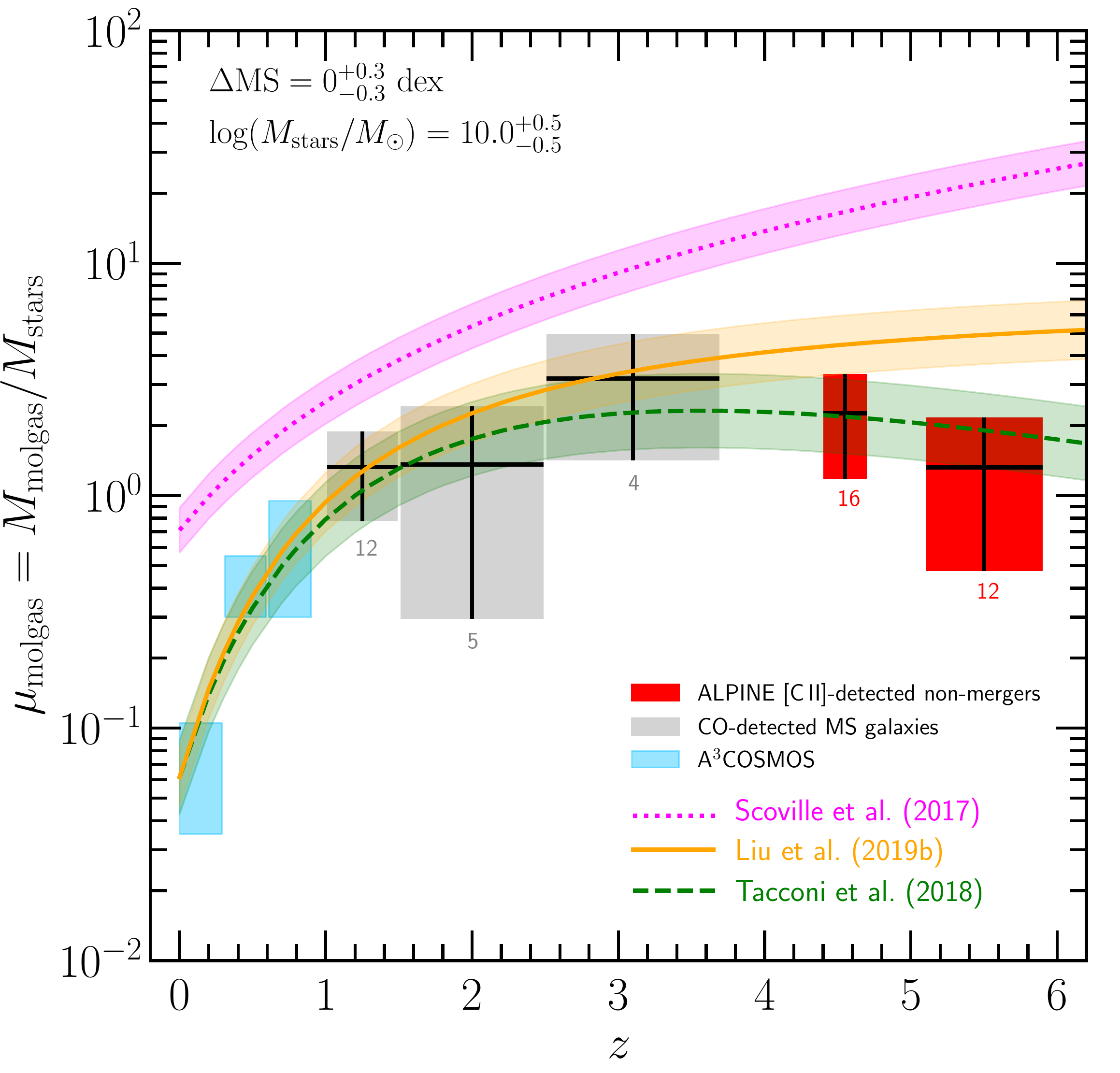}
\caption{Redshift evolution of the molecular depletion timescale ({\it top panels}) and the molecular gas mass to stellar mass ratio ({\it bottom panels}) of MS galaxies ($\Delta {\rm MS} = 0^{+0.3}_{-0.3}$~dex) in two stellar mass bins of $\log (M_{\rm stars}/M_{\odot}) = 8.4-9.5$ ({\it left panels}) and $\log (M_{\rm stars}/M_{\odot}) = 9.5-10.5$ ({\it right panels}). The red boxes show the respective $t_{\rm depl}$ and $\mu_{\rm molgas}$ means $\pm 1\,\sigma$ dispersion of the ALPINE [C\,{\sc ii}]-detected non-merger galaxies in redshift bins of $4.4<z<4.6$ and $5.1<z<5.9$. The grey boxes represent the CO-detected galaxies from our compilation in redshift bins of $1<z<1.5$, $1.5<z<2.5$, and $2.5<z<3.7$, and the blue boxes the A$^3$COSMOS galaxies at $0<z<1$ in $\Delta z=0.3$ bins. The number drawn below boxes gives the number of galaxies used to derive the mean and $1\,\sigma$ dispersion. For comparison, we show with violet dotted, orange solid, and green dashed lines the multi-functional $t_{\rm depl}$ and $\mu_{\rm molgas}$ best-fit functions of, respectively, \citet{scoville17}, \citet{liu19b}, and \citet{tacconi18}, calculated for $\Delta {\rm MS} = 0$~dex (the shaded areas define the $\Delta {\rm MS} = \pm 0.3$~dex range) and for fixed stellar masses of $\log (M_{\rm stars}/M_{\odot}) = 9.2$  ({\it left panels}) and $\log (M_{\rm stars}/M_{\odot}) = 10$ ({\it right panels}).}
\label{fig:tdepl-mugas-functions}
\end{figure*}
%
%-----------------------------------------------------------------------

There is a significant scatter, larger than 1~dex, among the $t_{\rm depl}$ measurements in all redshift bins, even though we only consider MS galaxies with $\Delta {\rm MS} = \pm 0.3$~dex around the MS parametrization of \citet{speagle14}. This scatter at a fixed redshift is believed to be a product of the multi-functional dependence of $t_{\rm depl}$ on many physical parameters, such as the offset from the MS, the star formation rate, the stellar mass, and possibly the environment \citep[e.g.,][]{dessauges15,scoville17,noble17,silverman18,tacconi18,tadaki19,liu19b}. Given the strong anti-correlation found between $t_{\rm depl}$ and the offset from the MS \citep{genzel15,dessauges15,tacconi18}, we still expect $t_{\rm depl}$ variations for galaxies on the MS while in their evolutionary process they are transiting up and down across the MS band \citep[e.g.,][]{sargent14,tacchella16}. The previously reported anti-correlation between $t_{\rm depl}$ and sSFR \citep{saintonge11,dessauges15} is also further supported by our galaxies at $z=4.4-5.9$ (Fig.~\ref{fig:tdepl}, middle panel). It highlights comparable timescales for gas consumption and stellar mass formation \citep{saintonge11,dessauges15}. We find a Spearman rank coefficient of $-0.49$ and $p$-value of $4.5\times 10^{-10}$ for the dependence of $t_{\rm depl}$ on sSFR when considering the MS SFGs at $z\sim 1-5.9$. The observed offset of ALPINE galaxies with respect to the $t_{\rm depl}$--sSFR relation of MS SFGs at $z=0$ and to a lesser extent to the relations at $z\sim 1$ and $z\sim 2$ agrees with the displacement of the $z=0$ relation along the sSFR-axis by factors derived from the redshift evolution of sSFR of MS SFGs out to redshifts of $z\sim 5$ \citep{speagle14}, although a less steep sSFR redshift evolution toward $z\sim 5$ than parametrised by \citet{speagle14} is suggested in line with the ${\rm sSFR}(z)$ results of \citet{khusanova20b}. With $t_{\rm depl}$ measurements achieved down to $M_{\rm stars} \sim 10^{8.4}~M_{\odot}$ for the ALPINE galaxies, we confirm, on the other hand, that for MS SFGs at $z\sim 1-5.9$ the $t_{\rm depl}$ dependence on $M_{\rm stars}$, if any, must be weak as shown in Fig.~\ref{fig:tdepl} (right panel). This supports that the linear Kennicutt-Schmidt relation established for local galaxies \citep{kennicutt98a} might hold up to $z\sim 5.9$ MS SFGs.

\citet{scoville17}, \citet{tacconi18}, and \citet{liu19b} performed, for their respective compilations of galaxies with $M_{\rm molgas}$ measurements, a multi-functional fitting to simultaneously quantify the underlying dependency of $t_{\rm depl}$ as products of power laws in redshift, $M_{\rm stars}$, and offset from the MS (as well as optical size in the case of \citet{tacconi18}, who ultimately found a negligible $t_{\rm depl}$ dependence on size). They used slightly different criteria in their fitting procedure, but assumed the same MS parametrization from \citet[][Eq.~(28)]{speagle14}\footnote{To be exact, \citet{scoville17} used a combination of MS parametrizations from \citet{speagle14} and \citet{lee15}, but this combination only affects SFGs with high $M_{\rm stars} \gtrsim 10^{10.5}~M_{\odot}$. Below this mass threshold, the \citet{speagle14} MS parametrization holds.}. Their respective best-fits yield different $t_{\rm depl}$ functional forms, which are compared in \citet{liu19b}. While the \citet{tacconi18} $t_{\rm depl}$ function was fitted with data covering only redshifts of $z\sim 0-3$, the \citet{liu19b} function does account for data at $z>3$, but restricted to MS SFGs with high $M_{\rm stars}$ ($M_{\rm stars} \sim 10^{11}~M_{\odot}$). The three fitted functions in fact lack constraints for MS low mass ($M_{\rm stars} \lesssim 10^{10}~M_{\odot}$) SFGs at $z>3$. These SFGs are particularly important, since, as shown by \citet{liu19b}, the largest differences between the three fitted $t_{\rm depl}$ functions are observed for MS SFGs at $z>4$ with $M_{\rm stars} < 10^{10}~M_{\odot}$. The ALPINE galaxies are precisely characterised by these physical properties and can therefore bring decisive constraints on the $t_{\rm depl}$ function. 

In Fig.~\ref{fig:tdepl-mugas-functions} (top panels) we show, similarly to \citet{liu19b}, the molecular gas depletion timescale as a function of redshift as predicted by the three $t_{\rm depl}$ best-fit functions for MS galaxies with $\Delta {\rm MS}$ ranging from $-0.3$~dex to $+0.3$~dex and stellar masses in two bins of $\log (M_{\rm stars}/M_{\odot}) = 9.2^{+0.3}_{-0.8}$ and $\log(M_{\rm stars}/M_{\odot}) = 10^{+0.5}_{-0.5}$. To compare the observations with the plotted best-fit functions, we bin the ALPINE galaxies in two redshift intervals of $4.4<z<4.6$ and $5.1<z<5.9$ (red boxes), and the CO-detected MS SFGs from our compilation (Sect.~\ref{sect:compilation}) in three redshift intervals of $1<z<1.5$, $1.5<z<2.5$, and $2.5<z<3.7$ (grey boxes). 
%and we show the respective mean and $1\,\sigma$ dispersion in each of these redshift bins. The blue boxes represent the mean and $1\,\sigma$ dispersion in $\Delta z=0.3$ bins
The blue boxes represent MS SFGs at $0<z<1$ from A$^3$COSMOS in $\Delta z=0.3$ bins \citep{liu19b}. The ALPINE galaxies exclude the $t_{\rm depl}$ best-fit function of \citet{liu19b} at $z\gtrsim 4.5$ in the two $M_{\rm stars}$ bins, but already in the redshift bin of $2.5<z<3.7$ we observe a deviation from this function in the $\log(M_{\rm stars}/M_{\odot}) = 10^{+0.5}_{-0.5}$ bin. On the other hand, both the \citet{scoville17} and \citet{tacconi18} $t_{\rm depl}$ functions agree with the ALPINE observations, 
%with maybe a slightly better agreement with the \citet{tacconi18} function if 
and this even if we consider the possible $\rm SFR_{IR}$ contribution for the ALPINE galaxies undetected in the FIR dust continuum (see Sect.~\ref{sect:observations}), which would lower the plotted $t_{\rm depl}$ means by a factor of 1.5 in the redshift bin of $4.4<z<4.6$ and less in the higher redshift bin. The discrepancy of the \citet{liu19b} best-fit function with the other two functions results from the strong anti-correlation they find between $t_{\rm depl}$ and $M_{\rm stars}$.
%the lower the stellar mass is, the longer the molecular gas depletion timescale is. 
This dependence of $t_{\rm depl}$ on $M_{\rm stars}$ is too strong for SFGs with $M_{\rm stars} < 10^{10.5}~M_{\odot}$ at $z\gtrsim 3$, but seems to be correct at the high $M_{\rm stars}$ end of $M_{\rm stars} \gtrsim 10^{11}~M_{\odot}$ where both the \citet{scoville17} and \citet{tacconi18} functions overestimate the $t_{\rm depl}$ measurements at $z\gtrsim 3$ \citep[see Fig.~12 in][]{liu19b}. We postpone the refitting of the functional form of $t_{\rm depl}$ by including ALPINE galaxies in order to determine the scaling relation of $t_{\rm depl}$ over a more complete $M_{\rm stars}$ and redshift range to a future paper.
%We plan to refit the functional form of $t_{\rm depl}$ by including the ALPINE galaxies in a subsequent paper in order to derive the correct scaling relation of $t_{\rm depl}$ over a complete stellar mass and redshift range.

%
%Dense PDR associated with molecular clouds
%total gas mass including diffuse neutral gas
%don't see continuous decrease 
%
%-----------------------------------------------------------------------

\subsection{Molecular gas fraction}
\label{sect:fmolgas}

\begin{figure}
\centering
\includegraphics[width=8.5cm,clip]{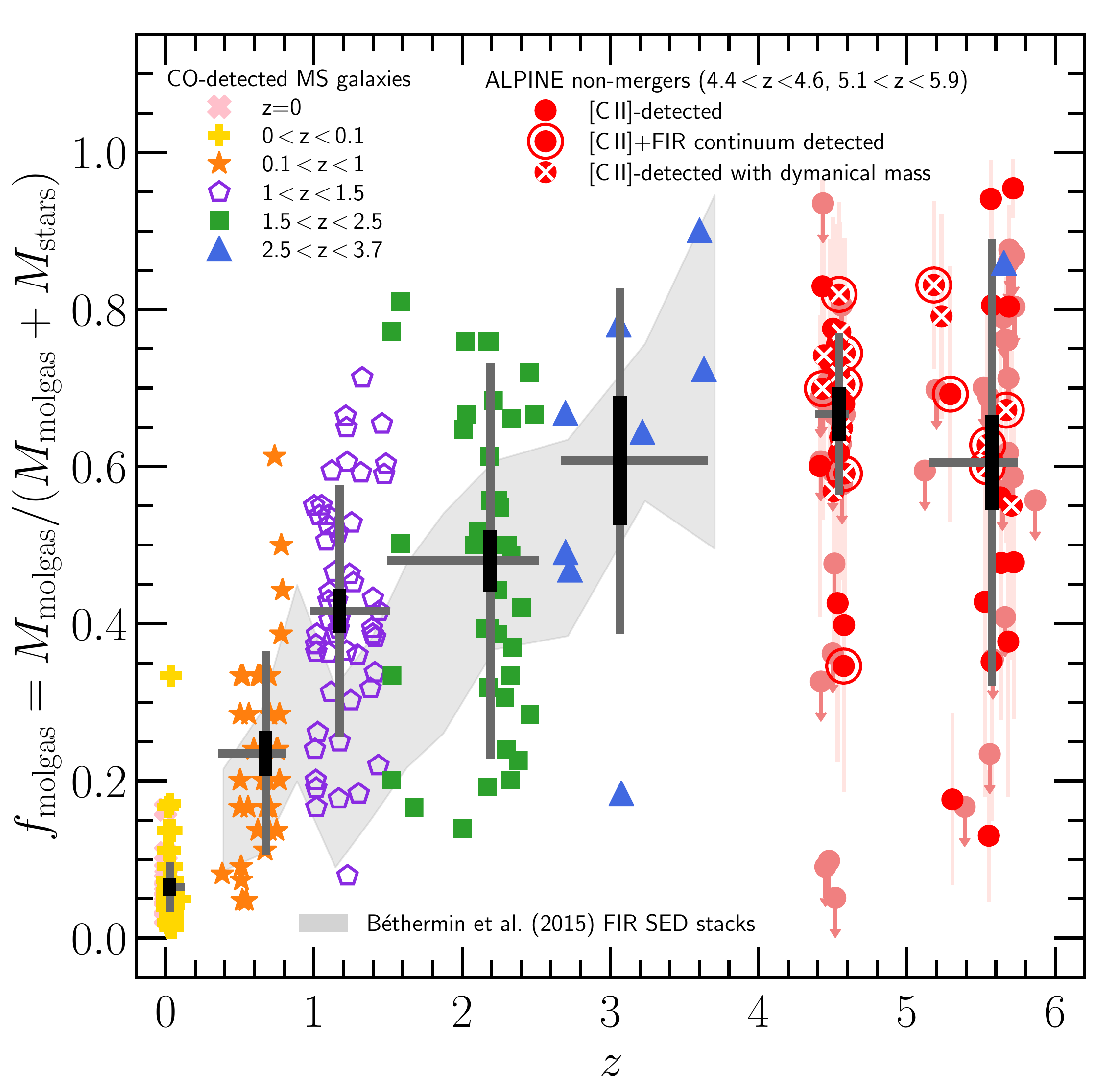}
\includegraphics[width=8.5cm,clip]{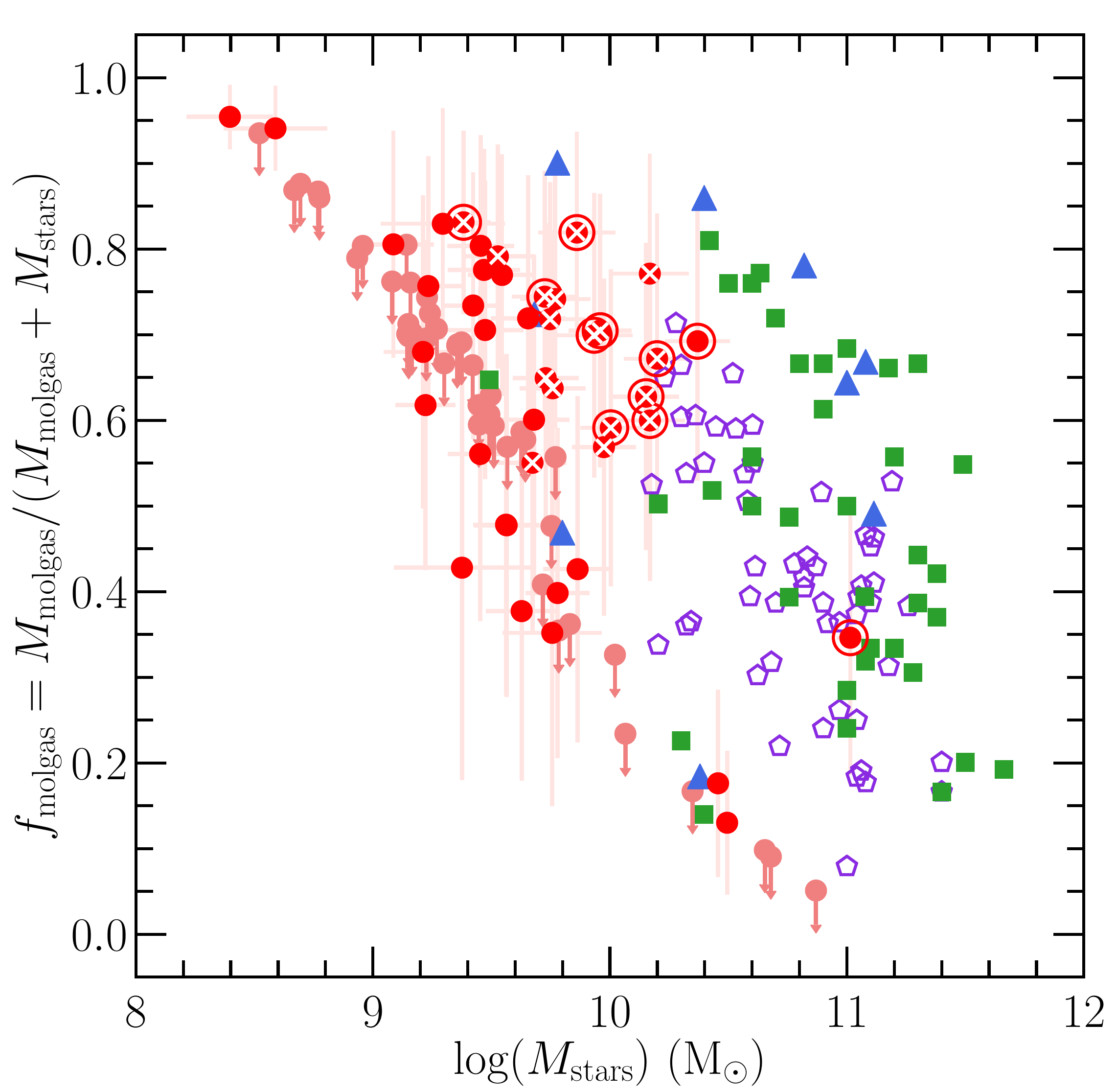}
\caption{Molecular gas fractions plotted for the same ALPINE galaxies (red circles) and CO-detected MS SFGs, with the same color-coding per redshift bin, as in Fig.~\ref{fig:tdepl}. {\it Top panel.} Molecular gas fractions shown as a function of redshift. The respective means, errors on the mean, and standard deviations per redshift bin are indicated by the black/grey big crosses. The light-grey shaded area corresponds to the molecular gas fractions obtained by \citet{bethermin15} from FIR SED stacks. The $f_{\rm molgas}$ means per redshift bin show a steep increase from $z=0$ to $z\sim 3.7$, followed by a flattening toward higher redshifts within the $1\,\sigma$ dispersion on the means.
{\it Bottom panel.} Molecular gas fractions, restricted to $z\sim 1-5.9$ SFGs, shown as a function of stellar mass. A strong dependence of $f_{\rm molgas}$ on $M_{\rm stars}$ is observed for CO-detected high-redshift galaxies and the ALPINE galaxies as well.}
\label{fig:fmolgas}
\end{figure}

In Fig.~\ref{fig:fmolgas} (top panel) we show the molecular gas fraction, defined as $f_{\rm molgas} =  M_{\rm molgas}/(M_{\rm molgas} + M_{\rm stars})$, as a function of redshift for the ALPINE [C\,{\sc ii}]-detected non-merger galaxies (red circles) and [C\,{\sc ii}]-non-detected galaxies (light-red arrows) in the redshift bins of $4.4<z<4.6$ and $5.1<z<5.9$, and for our compilation of CO-detected MS SFGs from the literature separated in the same 6 redshift bins as in Sect.~\ref{sect:tdepl} and Fig.~\ref{fig:tdepl} of $z=0$, $0<z<0.1$, $0.1<z<1$, $1<z<1.5$, $1.5<z<2.5$, and $2.5<z<3.7$, chosen in the way that the three bins between $z=0.1$ and $z=2.5$ contain a comparable number of galaxies (about 40).
%and our compilation of CO-detected main-sequence SFGs from the literature separated in 6 redshift bins of $z=0$ (black thin crosses), $0<z<0.1$ (cyan thick crosses), $0.1<z<1$ (yellow stars), $1<z<1.5$ (blue open pentagons), $1.5<z<2.5$ (grey squares), and $2.5<z<3.7$ plus $z=5.65$\footnote{The redshift of the \citet{pavesi19} CO-detected galaxy.} (green triangles), with their respective means, errors on the mean, and standard deviations (black/grey big crosses).
We then compute the respective means, errors on the mean, and standard deviations per redshift bin (black/grey big crosses). We show the ALPINE means obtained for the 44 [C\,{\sc ii}] detections (see Table~\ref{tab:tdeplfmolgas}). We also overplot the \citet{bethermin15} FIR SED stacks (light-grey shaded area). We observe a steep rise of $f_{\rm molgas}$ from $z=0$ to $z\sim 3.7$, in agreement with what has been previously reported \citep[e.g.,][]{dessauges17,scoville17,tacconi18,tacconi20}. With the ALPINE sample we probe, for the first time, the $f_{\rm molgas}$ evolution beyond $z\gtrsim 4.5$ in MS SFGs with a low median $M_{\rm stars}$ of $10^{9.7}~M_{\odot}$. Within the $1\,\sigma$ dispersion on the $f_{\rm molgas}$ means in the two redshift bins, we observe a flattening of $f_{\rm molgas}$ that reaches a mean value of $63\%\pm 3\%$ over $z=4.4-5.9$ (Table~\ref{tab:tdeplfmolgas}). 
%When taking the means per redshift bin at face value, the ALPINE galaxies even seem to show some hint of $f_{\rm molgas}$ decrease in the highest $5.1<z<5.9$ redshift bin, but statistically not significant.
The observed flattening is not subject to the assumptions that are needed to translate [C\,{\sc ii}] luminosities into molecular gas masses, since both $4.4<z<4.6$ and $5.1<z<5.9$ redshift bins are subject to those assumptions in the same way. When applying the survival analysis to take into account the ``secure'' $3\,\sigma$ upper limits of the ALPINE galaxies undetected in [C\,{\sc ii}] (Sect.~\ref{sect:tdepl}), the $f_{\rm molgas}$ means in the $4.4<z<4.6$ and $5.1<z<5.9$ redshift bins drop to $46\%\pm 5\%$ (Table~\ref{tab:tdeplfmolgas}). This strengthens the $f_{\rm molgas}$ flattening toward high redshifts, which is an important result, consistent with the evolutionary trend of a constant sSFR beyond $z\gtrsim 4$ obtained by several studies \citep[e.g.,][]{tasca15,khusanova20a}, including the sSFR derived from the obscured SFR measured in the ALPINE galaxies by stacking the FIR dust continuum maps in the redshift bins of $4.4<z<4.6$ and $5.1<z<5.9$ \citep{khusanova20b}. The finding that $f_{\rm molgas}$ and sSFR merely have a similar evolution with redshift is not a surprise, since $f_{\rm molgas}$ can be expressed as a function of $t_{\rm depl}$ and sSFR \citep{tacconi13}:
\begin{equation}
f_{\rm molgas} = \frac{1}{1+({\rm sSFR}~t_{\rm depl})^{-1}}.
\end{equation}
As a result, the $f_{\rm molgas}$ redshift evolution depends on the redshift evolution of both $t_{\rm depl}$ and sSFR. In the case of, on average, a weak change in $t_{\rm depl}$ of MS SFGs with redshift, which is what we observe in Fig.~\ref{fig:tdepl} (left panel), we globally have $f_{\rm molgas}(z) \propto {\rm sSFR}(z)$.

%
%-----------------------------------------------------------------------
%
%-----------------------------------------------------------------------

In the framework of the bathtub model, the $f_{\rm molgas}$ evolution with redshift reflects an interplay between cosmic inflow (supply of fresh gas onto galaxies) and gas consumption rates, modulo outflows. The mass accretion rate was shown to scale as $(1+z)^{2.25}$ \citep{dekel09}, therefore 
%and the gas consumption rate is given by how fast gas can be processed into stars or an outflow. Since $t_{\rm depl}^{-1}$ evolves with redshift slower than $(1+z)^{2.25}$ (see Sect.~\ref{sect:tdepl}), 
the gas supply rate drops faster with time than the gas consumption rate (see Sect.~\ref{sect:tdepl}). This explains why galaxies at sufficiently high redshifts begin to be gas-rich, but then $f_{\rm molgas}$ drops as the gas consumption rate catches up. The phase during which galaxies have an excess of gas, and hence are in non-equilibrium, will also depend on feedback, since outflows, by ejecting the gas out of galaxies, reduce the amount of gas that needs to be processed into stars and help to establish the equilibrium earlier on. A quick look at the $f_{\rm molgas}$ observations supports a gas excess until at most $z\sim 3$ (Fig.~\ref{fig:fmolgas}, top panel). This is much shorter in cosmic time than predicted by cosmological simulations of \citet{lagos15}, who report a drop of $f_{\rm molgas}$ only several Gyr later, by $z\sim 1$. Given the shallow $t_{\rm depl}$ evolution with redshift, outflows must play an important role in blowing out part of the infalling gas at $z\gtrsim 3$. This is supported by signatures of star formation-driven outflows in stacks of [C\,{\sc ii}] spectra and [C\,{\sc ii}] moment-zero maps and stacks of rest-frame UV spectra of the ALPINE higher SFR ($\gtrsim 25~M_{\odot}$~yr$^{-1}$) galaxies \citep{ginolfi19,faisst19}, but also observed in a few individual ALPINE objects with [C\,{\sc ii}] halos \citep{fujimoto19,ginolfi20}. Observational evidence of star formation-driven outflows in SFGs at $z\lesssim 5-6$ was also reported in other studies \citep[e.g.,][]{sugahara19,rubin14,talia17}.
%the contribution of active galactic nuclei feedback in MS SFGs being so far unconstrained
%the detection of [C\,{\sc ii}] outflows in the ALPINE galaxies at higher SFRs \citep{ginolfi19} .

%
%-----------------------------------------------------------------------
\begin{figure*}
\centering
\includegraphics[width=8.5cm,clip]{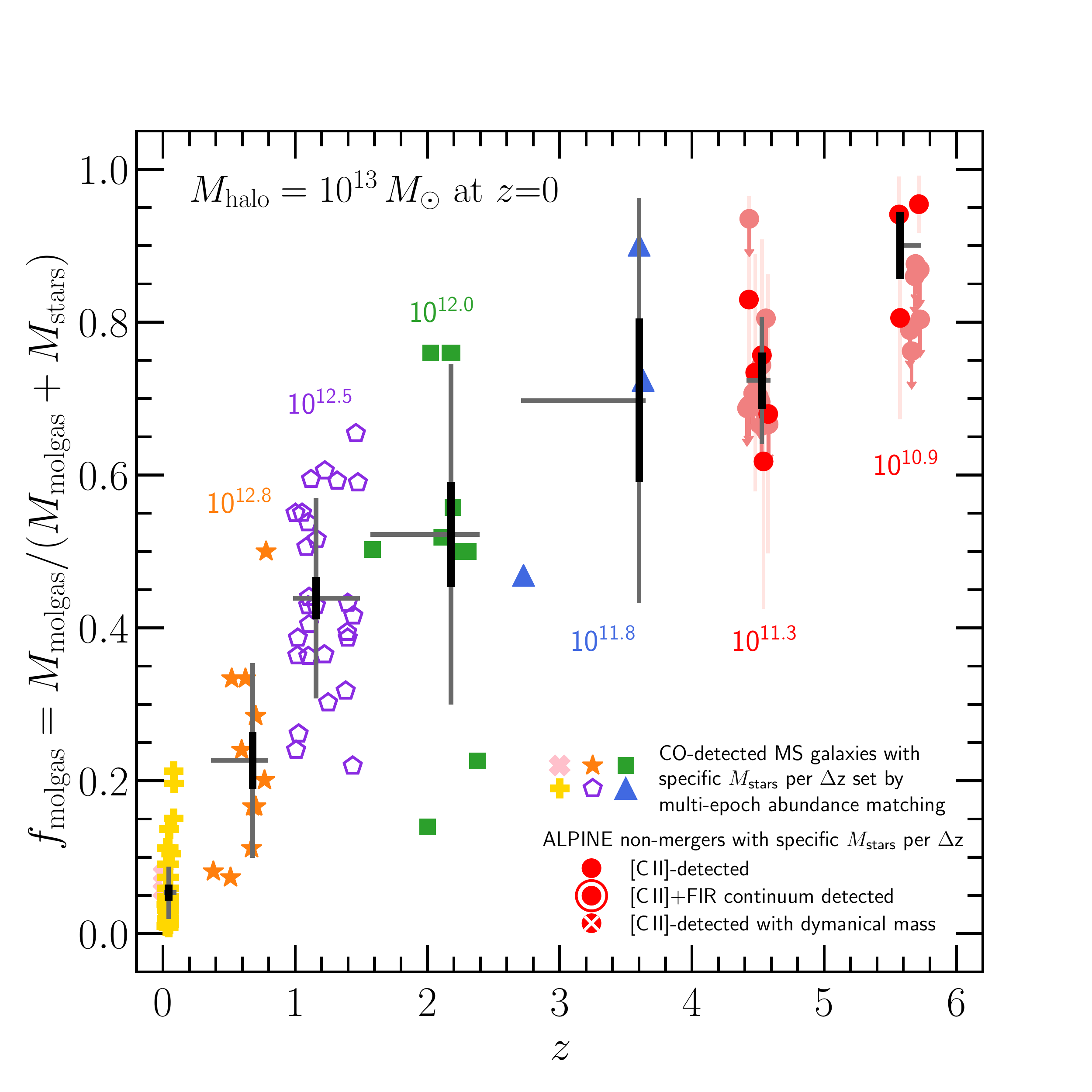} \hspace{0.2cm}
\includegraphics[width=8.5cm,clip]{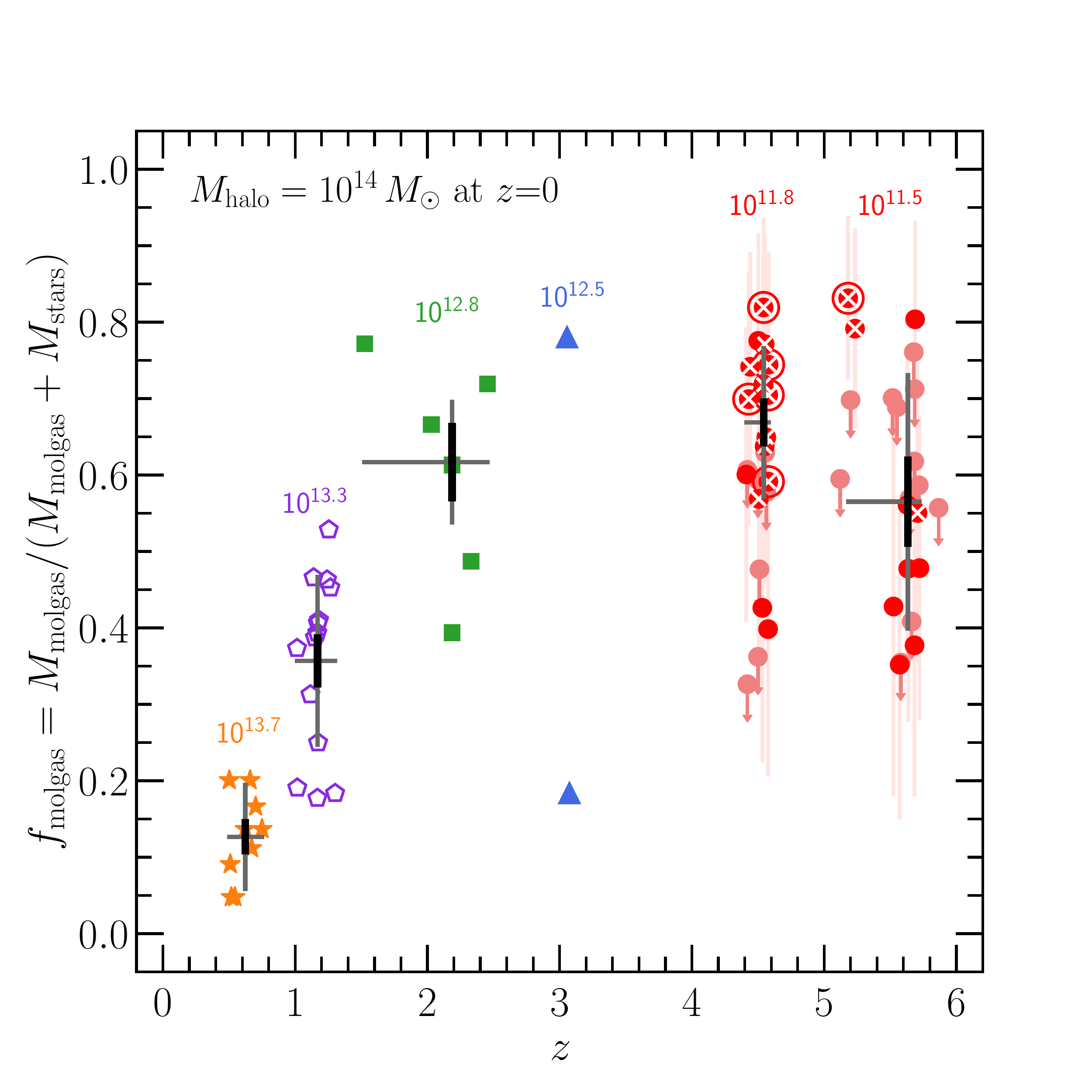}
\caption{Evolution of the molecular gas fraction with redshift plotted for the same ALPINE galaxies (red circles) and CO-detected MS SFGs, with the same color-coding per redshift bin, as in Fig.~\ref{fig:tdepl}, but restricted to the $z > 0-5.9$ progenitors of, respectively, Milky Way-like galaxies at $z=0$ with stellar masses in the range of $\sim 10^{10.8}-10^{11.2}~M_{\odot}$ for a halo mass of $10^{13}~M_{\odot}$ ({\it left panel}), and more massive $z=0$ galaxies with $M_{\rm stars} \sim 10^{11.4}-10^{11.7}~M_{\odot}$ for a halo mass of $10^{14}~M_{\odot}$ ({\it right panel}). We consider the progenitors' $M_{\rm stars}$ as a function of redshift listed in Table~\ref{tab:abundance-matching}, obtained from the multi-epoch abundance matching predictions of \citet{behroozi19}. The number drawn in each redshift bin corresponds to $M_{\rm halo}$ at this epoch. A different molecular gas fraction evolution from $z=5.9$ to $z=0$ is observed for the respective progenitors of the $10^{13}~M_{\odot}$ and $10^{14}~M_{\odot}$ halo mass galaxies at $z=0$.}
\label{fig:fmolgas-MEAM}
\end{figure*}
%
%-----------------------------------------------------------------------

%While we may see in the $f_{\rm molgas}$ flattening toward high redshifts a saturation of the molecular gas mass with respect to the stellar mass, this is not really true when considering galaxies individually, since 
While we observe an overall flattening of $f_{\rm molgas}$ toward high redshifts, some individual galaxies appear to depart from this average trend considerably: the scatter in $f_{\rm molgas}$ among ALPINE MS SFGs ranges from $\sim 15$\% to $\sim 95$\%. A significant scatter among $f_{\rm molgas}$ measurements is observed in all redshift bins (although the scatter is particularly large at $5.1<z<5.9$). The tight correlation between $f_{\rm molgas}$ and offset from the MS, reported even for MS SFGs lying within the $\pm 0.3$~dex dispersion of the MS \citep{tacconi13,dessauges15,genzel15,saintonge16}, certainly contributes to this $f_{\rm molgas}$ scatter per redshift bin. In addition to that, there is a strong dependence of $f_{\rm molgas}$ on $M_{\rm stars}$ as shown in the bottom panel of Fig.~\ref{fig:fmolgas}, previously found for local and $z\lesssim 3$ MS SFGs \citep[e.g.,][]{saintonge11,tacconi13,tacconi18,tacconi20,dessauges15,schinnerer16,scoville17}, and now assessed for the ALPINE $z=4.4-5.9$ galaxies (Spearman rank coefficient of $-0.50$ and $p$-value of $7.0\times 10^{-4}$). The observed steep drop in $f_{\rm molgas}$ with increasing $M_{\rm stars}$ is expected from the gas conversion into stars, and is predicted by semi-analytical simulations developed in the framework of the bathtub model, as well as cosmological hydrodynamic simulations, for both local and high-redshift galaxies \citep{bouche10,dave11,dave17}. This behaviour was proposed to be consistent with the ``downsizing'' scenario, where at fixed redshift massive galaxies have lower $f_{\rm molgas}$ because they consumed their fuel of star formation earlier than less massive galaxies that still have large fractions of gas \citep{bouche10,santini14,dessauges15,scoville17}. We find that the significant $f_{\rm molgas}$ scatter of ALPINE galaxies must be mostly driven by the large range of $M_{\rm stars}$, from $10^{8.4}~M_{\odot}$ to $10^{11}~M_{\odot}$, they encompass. 
%At fixed stellar mass the currently available data do not show evidence for the predicted shift in $f_{\rm molgas}$ toward higher values at higher redshifts. 
%The location of the upturn of $f_{\rm molgas}$ at the low-mass end ($M_{\rm stars} \sim 10^9~M_{\odot}$) is shown to be dependent on whether outflows are considered or not.

Similarly to the multi-functional fitting performed by \citet{scoville17}, \citet{tacconi18}, and \citet{liu19b} 
%to quantify the dependency of the molecular gas depletion timescale, 
for $t_{\rm depl}$, we show in the bottom panels of Fig.~\ref{fig:tdepl-mugas-functions} the best-fit functions obtained for the molecular gas ratio, $\mu_{\rm molgas} = M_{\rm molgas}/M_{\rm stars}$, as a function of redshift for MS galaxies with $\Delta {\rm MS}$ ranging from $-0.3$~dex to $+0.3$~dex and stellar masses in two bins of $\log (M_{\rm stars}/M_{\odot}) = 9.2^{+0.3}_{-0.8}$ and $\log(M_{\rm stars}/M_{\odot}) = 10^{+0.5}_{-0.5}$. We consider these two $M_{\rm stars}$ bins, because it is at these $M_{\rm stars}$ that the larger differences between the three fitted $\mu_{\rm molgas}$ functions are found. To compare the observations with the plotted best-fit functions, we bin again the ALPINE galaxies in two redshift intervals of $4.4<z<4.6$ and $5.1<z<5.9$ (red boxes), and the CO-detected MS SFGs from our compilation (Sect.~\ref{sect:compilation}) in three redshift intervals of $1<z<1.5$, $1.5<z<2.5$, and $2.5<z<3.7$ (grey boxes). 
%and plot the respective mean and $1\,\sigma$ dispersion in each of these redshift bins. The blue boxes represent the mean and $1\,\sigma$ dispersion in $\Delta z=0.3$ bins of
The blue boxes represent MS SFGs at $0<z<1$ from A$^3$COSMOS in $\Delta z=0.3$ bins \citep{liu19b}. Our data favour the \citet{tacconi18} best-fit function, given the comparable decrease of the predicted and measured $\mu_{\rm molgas}$ at $z=4.4-5.9$, and the $t_{\rm depl}$ results discussed in Sect.~\ref{sect:tdepl}. This function also provides a good fit to the $\mu_{\rm molgas}$ redshift evolution of massive MS SFGs \citep[see Fig.~13 in][]{liu19b}. On the other hand, the \citet{scoville17} function overestimates the molecular gas ratios of MS SFGs in both $M_{\rm stars}$ bins considered.

%
%-----------------------------------------------------------------------

\subsection{Molecular gas fraction over cosmic time}
\label{sect:fmolgas-evolution}
%radiation pressure preventing infalling gas.

As stressed by \citet{wiklind19}, to probe the true evolution of galaxy properties over cosmic time, galaxies need to be carefully selected in a way which correctly connects the progenitors at high redshifts with their descendants at $z=0$. A possible selection method is to use the multi-epoch abundance matching, which links as a function of redshift the growth of central dark matter halos, as derived from numerical simulations, with the growth of stellar content constrained from observations of the $M_{\rm stars}$ function \citep{behroozi13,behroozi19,moster13,moster18}. The redshift evolution of the resulting stellar-to-halo mass relation is driven by gas accretion, star formation, feedback (leading to stellar mass loss), and eventually merging processes. 

Following the work by \citet[][top right panel of Fig.~18]{behroozi19}, we use the evolutionary corridors they computed, in the $M_{\rm stars}$ versus redshift plane, for the stellar mass histories of progenitors of $z=0$ galaxies with a given $M_{\rm stars}$ range. 
%the regions they define in the stellar mass versus redshift space, where progenitors of $z=0$ galaxies with a given stellar mass range are expected. 
The ALPINE [C\,{\sc ii}]-detected non-merger galaxies with their $M_{\rm stars}$ appear to be the progenitors, at $z\sim 4.5$ and $z\sim 5.5$, of Milky Way-like galaxies at $z=0$ with $M_{\rm stars}$ in the range of $\sim 10^{10.8}~M_{\odot}$ and $10^{11.2}~M_{\odot}$ and of more massive $z=0$ galaxies with $M_{\rm stars} \sim 10^{11.4}-10^{11.7}~M_{\odot}$. The range of $M_{\rm stars}$ of these $z=0$ descendants with halo masses of $M_{\rm halo} = 10^{13}~M_{\odot}$ and $10^{14}~M_{\odot}$ at $z=0$, respectively, were carefully chosen such that their respective stellar mass histories do not overlap in the $M_{\rm stars}$--$z$ plane. In Table~\ref{tab:abundance-matching} we list the respective stellar mass histories. We then select, in our compilation of CO-detected MS SFGs and ALPINE galaxies, progenitors with the specific $M_{\rm stars}$ over cosmic time from $z>0$ to $z=5.9$.
%of progenitors' stellar masses as a function of redshift of these galaxies with halo masses of $M_{\rm halo} = 10^{13}~M_{\odot}$ and $10^{14}~M_{\odot}$ at $z=0$, respectively, are chosen in a way their stellar mass histories do not overlap in the $M_{\rm stars}$--$z$ space. They are listed in Table~\ref{tab:abundance-matching}. We then select, in our compilation of CO-detected main-sequence SFGs and ALPINE galaxies, progenitors with the right stellar masses over $z>0-5.9$.
%across the redshift range of $z>0-5.9$, representative of present-day galaxies with stellar masses in these two mass bins (see Table~\ref{tab:abundance-matching}).

%
%-----------------------------------------------------------------------
\begin{table}
\caption{Stellar mass histories from multi-epoch abundance matching predictions of \citet{behroozi19}}             
\label{tab:abundance-matching} 
\centering 
\begin{tabular}{c r@{ $\times$ }l r@{ $\times$ }l} 
\hline\hline
 & \multicolumn{2}{l}{$M_{\rm halo} = 10^{13}~M_{\odot}$ at $z=0$} & \multicolumn{2}{l}{$M_{\rm halo} = 10^{14}~M_{\odot}$ at $z=0$} \\
$\langle z\rangle$ & \multicolumn{2}{l}{$M_{\rm stars}$ range ($M_{\odot}$)} & \multicolumn{2}{l}{$M_{\rm stars}$ range ($M_{\odot}$)} \\
\hline  
0    &  $(5.9-16)$&$10^{10}$ & $(2.5-5.0)$&$10^{11}$ \\
0.7  &  $(3.4-12)$&$10^{10}$ & $(1.2-2.7)$&$10^{11}$ \\
1.2  &  $(2.2-10)$&$10^{10}$ & $(1.0-1.8)$&$10^{11}$ \\
2.2  &  $(3.0-43)$&$10^9$    &  $(4.3-10)$&$10^{10}$ \\
3.0  & $(8.0-180)$&$10^8$    & $(1.8-8.0)$&$10^{10}$ \\
4.5  &  $(1.0-29)$&$10^8$    &  $(2.9-27)$&$10^9$ \\
5.5  & $(3.2-130)$&$10^7$    & $(1.3-8.0)$&$10^9$ \\
\hline     
\end{tabular}
\end{table}
%
%-----------------------------------------------------------------------

As shown in Fig.~\ref{fig:fmolgas-MEAM}, we find a different evolution of the molecular gas fraction from $z=5.9$ to the present day for progenitors of Milky Way-analogues and for more massive $z=0$ galaxies.
%With the selected galaxies, we may, for the first time, study the redshift evolution of the molecular gas fraction of progenitors of Milky Way-like galaxies from $z=5.9$ until now. The results are shown in Fig.~\ref{fig:fmolgas-MEAM}. A different evolution of the molecular gas fraction with redshift is observed for progenitors of Milky Way analogues and more massive $z=0$ galaxies. The former 
Progenitors of Milky Way-like galaxies follow a monotonic decrease of $f_{\rm molgas}$ with cosmic time, which steepens at $z\lesssim 1$. However, this result relies on only three ALPINE $f_{\rm molgas}$ measurements in the redshift bin of $5.1<z<5.9$, which show galaxies dominated by gas with a mean $f_{\rm molgas}$ as high as $90\%\pm 4\%$. A larger sample of low $M_{\rm stars}$ galaxies in this redshift bin is necessary to confirm the currently observed monotonic decrease. Progenitors of the more massive $z=0$ galaxies show, on the other hand, a steep $f_{\rm molgas}$ decline at $z\lesssim 2$, which is preceded by a flat $f_{\rm molgas}$ evolution at higher redshifts, with a mean value of $63\%\pm 3\%$ at $z=4.4-5.9$, although some hint of an $f_{\rm molgas}$ rise from $z\sim 5.5$ to $z\sim 4.5$ exits. How can we explain these different $f_{\rm molgas}$ evolutions with cosmic time?

As discussed in Sect.~\ref{sect:fmolgas}, galaxies are believed to be supplied with cold gas by cosmic accretion flows. This accreted gas can then be used for the $M_{\rm stars}$ build-up of galaxies, if not partly expelled by outflows from the galaxy. 
%The drop of the molecular gas fraction marks the transition from a gas-dominated phase into a gas consumption phase, which will happen more or less early in time depending on feedback effects. 
\citet{ginolfi19} showed evidence of star formation-driven outflows in the [C\,{\sc ii}] emission stacks of the ALPINE galaxies with SFR higher than the median SFR of the APLINE sample (${\rm SFR} > 25~M_{\odot}~\rm yr^{-1}$). These higher SFR galaxies are also the more massive ones, due to their placement on the MS \citep{faisst19}. As a result, the observed outflows could contribute more to moderate the gas content available for star formation in the massive ALPINE galaxies, progenitors of $10^{14}~M_{\odot}$ halo mass galaxies at $z=0$, and could explain their flat $f_{\rm molgas}$ evolution from $z=5.9$ to $z=4.4$ (Fig.~\ref{fig:fmolgas-MEAM}, right panel) given also the induced quenching of star formation yielding a temporarily decrease of the gas consumption rate. This scenario matches with the non-detection of star formation-driven outflows in the less star-forming (and therefore, on average, less massive) ALPINE galaxies \citep{ginolfi19}, progenitors at $z=4.4-5.9$ of Milky Way-like galaxies, and hence with the observed steady decrease of their $f_{\rm molgas}$ from $z=5.9$ to $z=4.4$ (Fig.~\ref{fig:fmolgas-MEAM}, left panel). 

Moreover, it has also been suggested by simulations that for very massive dark matter halos 
%more massive than $M_{\rm halo} \sim 10^{13}~M_{\odot}$, 
the gas supply starts to shut off and prevents star formation 
\citep{dekel06,keres09,bouche10}. This is due to the fact that as the halo grows larger, it reaches the threshold for virial shock heating ($M_{\rm shock} \lesssim 10^{12}~M_{\odot}$) and, consequently, the infalling cold gas shock heats up close to the virial temperature \citep{dekel09}. 
%unless the gas penetrates into the hot dark matter halos along narrow streams where the cooling remains efficient . But still, the cold gas supply by streams is expected to drop in very massive halos at high redshift. The ALPINE galaxies 
Our data suggest that this might be happening at $z\sim 5$ in $\sim 10^{11.5-11.8}~M_{\odot}$ halos, the progenitors of $z=0$ halos of $10^{14}~M_{\odot}$ (see Fig.~\ref{fig:fmolgas-MEAM}, right panel). This indirectly implies that these massive galaxies must be mature by $z\sim 5$ and probably quench earlier than lower mass galaxies. Comparing right and left panels of Fig.~\ref{fig:fmolgas-MEAM}, we observe that the $f_{\rm molgas}$ means of massive galaxies are smaller by $\sim 10$\% (on absolute scale) in the redshift bins of $1<z<1.5$ and $0.1<z<1$ than those of lower mass galaxies.

On the other hand, massive progenitors have to grow a lot in the past to build up their large $M_{\rm stars}$. But, is there still enough cold gas for them to grow sufficiently quickly to reach these $M_{\rm stars}$ by $z\sim 5$, if we advocate that the cold gas accreted on galaxies is either removed by outflows or reduced because of the suppression of the cosmic accretion flows? The constant $f_{\rm molgas}$ observed between $z=5.9$ and $z=4.4$, with even a possibly lower $f_{\rm molgas}$ value in the $5.1<z<5.9$ redshift bin, may also be the result of an efficient star formation in these massive SFGs, such that the infalling gas is rapidly converted into stars. Some evidence for a higher SFE in massive ALPINE galaxies is shown in Fig.~\ref{fig:tdepl} (right panel), where, when considering the ALPINE galaxies only, we see a trend for an anti-correlation between $t_{\rm depl}$ and $M_{\rm stars}$ (Spearman rank coefficient of $-0.33$ and $p$-value of 0.0016). \citet{liu19b} reported such an anti-correlation, but for the whole sample of SFGs from $z\sim 6$ to $z=0$. We find, on the contrary, that this anti-correlation is not present for MS SFGs at $z\lesssim 3.7$ \citep[see also][]{dessauges15,tacconi18}. We rather argue that the $t_{\rm depl}$ dependence on $M_{\rm stars}$ might change across the cosmic time, from a negative slope at high redshift to the positive slope that is observed for local galaxies \citep[e.g.,][]{saintonge17}. 
%We postpone a detailed analysis of the redshift evolution of the tdepl-Mstar relation to a future paper.

%Finally, the possibly lower gas fraction in the massive ALPINE galaxies at $z=5.1-5.9$ (Fig.~\ref{fig:fmolgas-MEAM}, right panel) requires to invoke either the presence of particularly more virulent outflows at these redshifts, or the onset at $z\sim 4.5$ of a more efficient mass accretion rate. However, the impact of this latter should also be visible in the $f_{\rm molgas}$ evolution of the less massive Milky Way progenitors at $z=4.4-4.6$, which is not the case.

%
%-----------------------------------------------------------------------

\section{Summary and conclusions}
\label{sect:conclusions}

%We took advantage of the 75 [C\,{\sc ii}]~158~$\mu$m detections from the ALPINE survey of main-sequence star-forming galaxies at $z=4.4-5.9$ \citep{lefevre19,faisst19,bethermin19} and the recently proposed correlation between [C\,{\sc ii}] luminosity and molecular gas mass by \citet{zanella18}, to estimate the galactic molecular gas content at these very high redshifts. Excluding the ALPINE [C\,{\sc ii}]-detected galaxies classified as mergers according to our morpho-kinematic visual classification \citep{lefevre19}, we obtain 44 molecular gas mass estimates of galaxies with a median stellar mass of $10^{9.7}~M_{\odot}$. 
We use observations from the ALPINE [C\,{\sc ii}]~158~$\mu$m survey of UV-selected star-forming galaxies at $z=4.4-5.9$ \citep{lefevre19,faisst19,bethermin19} and the correlation between [C\,{\sc ii}] luminosity and molecular gas mass recently proposed by \citet{zanella18}, to obtain $M^{\rm CII}_{\rm molgas}$ estimates of 44 [C\,{\sc ii}]-detected MS SFGs with a median $M_{\rm stars}$ of $10^{9.7}~M_{\odot}$ (Fig.~\ref{fig:Mmolgas-distributions}). 
%when excluding objects classified as mergers according to our morpho-kinematic visual classification \citep{lefevre19}. 
Prior our work, measurements of $M_{\rm molgas}$ at $z>4$ have been derived for 24 MS SFGs from A$^3$COSMOS based on the thermal dust FIR continuum emission \citep{liu19b}, all having $M_{\rm stars}$ one order of magnitude larger than those of ALPINE galaxies. And only two CO-derived molecular gas masses have been reported in MS SFGs at $z>5$ \citep{pavesi19,dodorico18}.

The [C\,{\sc ii}] luminosity appears to be an effective tracer of the gas mass of ALPINE galaxies with a $1\,\sigma$ uncertainty of $0.3$~dex. This level of error is assessed by comparing molecular gas mass estimates based on [C\,{\sc ii}] luminosity, rest-frame 158~$\mu$m continuum emission (measured for 11 galaxies; Fig.~\ref{fig:Mmolgas-Scoville}), and dynamical mass (determined for 17 galaxies with size measurements; Fig.~\ref{fig:Mmolgas-Dynamical}). The agreement with the dynamical mass, essentially probing the baryonic gas mass after removing $M_{\rm stars}$ (the relative dark matter contribution in the internal regions of galaxies is expected to be low), supports that $L_{\rm CII}$ likely traces the total gas mass including the molecular, atomic, and ionised gas phases, and hence possibly overestimates the true $M_{\rm molgas}$ unless the atomic and ionised gas masses are negligible as often assumed at high redshift. 

Accurate $M_{\rm stars}$ and SFR were obtained for the ALPINE galaxies from existing ancillary UV to IR data \citep{faisst19}. Together with $M_{\rm molgas}$, we derive fundamental physical quantities, which are the molecular gas depletion timescale and the molecular gas fraction, and explore scaling-relations between these physical quantities. To put the ALPINE galaxies in a global context, 
%we compare their molecular gas depletion timescales and molecular gas fractions with values obtained 
we build a comparison sample of MS SFGs between $z=0$ and $z\sim3.7$ with molecular gas mass measurements inferred from CO luminosities reported in the literature. %This comparison sample represents an exhaustive compilation of CO measurements for main-sequence SFGs at $z>1$. 
%built on the compilation by \citet{dessauges15,dessauges17}. 
Our main results can be summarized as follows:
\begin{itemize}
\item The ALPINE sample enables us to explore the $t_{\rm depl}$ evolution beyond $z\gtrsim 4.5$ for a statistically significant number of MS SFGs. We observe a continuous decline of $t_{\rm depl}$ from $z=0$ to $z\sim 5.9$, reaching a mean value of $(4.6\pm 0.8)\times 10^8$~yr at $z=5.1-5.9$, 
%with redshift with no break at $z> 4.4$ and up to $z=5.9$, 
which confirms a $t_{\rm depl}$ redshift evolution with a slope shallower than $(1+z)^{-1.5}$ that is predicted in the framework of the bathtub model (Fig.~\ref{fig:tdepl}, left panel). This suggests that MS SFGs at $z\gtrsim 4.5$ are not considerably more efficient in forming stars, only a factor of $2-3$ with respect to present-day galaxies, or undergo a significant increase in their $f_{\rm molgas}$. The large $t_{\rm depl}$ scatter of more than 1~dex per redshift bin is attributed to the multi-functional dependence of $t_{\rm depl}$ on various physical parameters, such as their offset from the MS, star formation rate, and stellar mass. With the ALPINE galaxies probing low-to-moderate $M_{\rm stars}$ with a median of $10^{9.7}~M_{\odot}$, we find that MS SFGs at $z\sim 1-5.9$ show no $t_{\rm depl}$ dependence on $M_{\rm stars}$ (Fig.~\ref{fig:tdepl}, right panel). %This supports that the local Kennicutt-Schmidt relation might hold up to $z\sim 5.9$.
%As a consequence, the multi-functional $t_{\rm depl}$ best-fit function of \citet{liu19b}, relying on a strong anti-correlation between $t_{\rm depl}$ and stellar mass, does not match the ALPINE $t_{\rm depl}$ measurements (Fig.~\ref{fig:tdepl-mugas-functions}, top panels), while it is the function that best reproduces $t_{\rm depl}$ values of galaxies with very high stellar masses of $M_{\rm stars} \gtrsim 10^{11}~M_{\odot}$ at $z\gtrsim 3$ \citep{liu19b}.

\item We confirm that the steep rise of $f_{\rm molgas}$ is confined between $z=0$ and $z\sim 3.7$. At higher redshift, as shown by ALPINE galaxies, the $f_{\rm molgas}$ evolution flattens and reaches a mean value as high as $63\%\pm 3\%$ over $z=4.4-5.9$ (Fig.~\ref{fig:fmolgas}, top panel). The $f_{\rm molgas}$ flattening is consistent with the sSFR redshift evolution, which also flattens beyond $z\gtrsim 4$ according to several studies, including the results on the obscured SFR measured in the ALPINE galaxies by stacking the FIR dust continuum maps \citep{khusanova20b}. The redshift, at which the turnover in the $f_{\rm molgas}$ evolution happens, depends on when the gas consumption rate catches up to the mass accretion rate. 
%that was shown to scale as $(1+z)^{2.25}$ \citep{dekel09}. 
We observe a gas excess until at most $z\sim 3$, which, given the slow $t_{\rm depl}$ evolution with redshift, suggests that outflows may play an important role in blowing out part of the infalling gas at $z\gtrsim 4$. We attribute the large scatter in $f_{\rm molgas}$ of ALPINE galaxies, from $\sim 10$\% to $\sim 95$\%, to mostly $M_{\rm stars}$ (Fig.~\ref{fig:fmolgas}, lower panel), given the strong dependence of $f_{\rm molgas}$ observed on $M_{\rm stars}$ (Fig.~\ref{fig:fmolgas}, bottom panel) that reflects an important ``downsizing'' effect from massive to low-mass galaxies at $z=4.4-5.9$.
%and we find that the multi-functional best-fit function of the molecular gas to stellar mass ratio of \citet{tacconi18} is favoured, while the one of \citet{scoville17} is excluded (Fig.~\ref{fig:tdepl-mugas-functions}, bottom panels). As a result, the \citet{tacconi18} function accommodates well both the depletion timescale and molecular gas to stellar mass ratio dependence on redshift of main-sequence SFGs with low and high stellar masses, except the very massive ones with $M_{\rm stars} \sim 10^{11.6}~M_{\odot}$.

\item The [C\,{\sc ii}]-detected ALPINE galaxies at $z=4.4-5.9$ and the currently available compilation of $f_{\rm molgas}$ measurements in CO-detected MS SFGs at $z\sim 0-3.7$ enables us, for the first time, to probe the $f_{\rm molgas}$ evolution over cosmic time, from $z=5.9$ to the present day, of the progenitors of $z=0$ Milky Way-like galaxies with $M_{\rm stars}$ in the range of $\sim 10^{10.8}-10^{11.2}~M_{\odot}$ for a  halo mass of $10^{13}~M_{\odot}$ and of more massive $z=0$ galaxies with $M_{\rm stars}\sim 10^{11.4}-10^{11.7}~M_{\odot}$ for a halo mass of $10^{14}~M_{\odot}$. We use the multi-epoch abundance matching predictions of \citet{behroozi19} to select progenitors' $M_{\rm stars}$ as a function of redshift (Table~\ref{tab:abundance-matching}). We observe a different redshift evolution of $f_{\rm molgas}$ for the respective progenitors of galaxy halos of different masses (Fig.~\ref{fig:fmolgas-MEAM}), where lower mass halos ($M_{\rm halo}=10^{13}~M_{\odot}$ at $z=0$) 
%progenitors of Milky Way-analogues and more massive $z=0$ galaxies (Fig.~\ref{fig:fmolgas-MEAM}): the former ..., and the latter ...
show a monotonic decrease of $f_{\rm molgas}$ with cosmic time and the higher mass halos ($M_{\rm halo}=10^{14}~M_{\odot}$ at $z=0$) show, on average, a flat $f_{\rm molgas}$ until $z\sim 2$ 
%with a possible trend of an $f_{\rm molgas}$ rise from $z\sim 5.5$ to $z\sim 4.5$, 
followed by a steep decrease at $z\lesssim 2$. This difference, if confirmed, likely reveals important changes in the physical conditions of MS SFGs at $z\sim 5$ for a specific halo mass threshold. 
%of their descendants at $z=0$. 
To explain the flat $f_{\rm molgas}$ evolution from $z=5.9$ to $z=4.4$ of progenitors of $z=0$ halos with $10^{14}~M_{\odot}$ masses, we discuss the effect of possibly stronger star formation-driven outflows in these more massive ALPINE galaxies as observed by \citet[][and also \citeauthor{fujimoto19} \citeyear{fujimoto19}, \citeauthor{faisst19} \citeyear{faisst19}]{ginolfi19},
%for the ALPINE galaxies with ${\rm SFR} > 25~M_{\odot}~\rm yr^{-1}$, 
which could remove their accreted cold gas and temporarily quench star formation. Simulations also suggest that for very massive dark matter halos, the gas supply starts to shut off. 
%and prevents star formation, 
According to the $f_{\rm molgas}$ redshift evolution observed for the ALPINE galaxies, 
%having for $z=0$ descendants $10^{14}~M_{\odot}$ halos 
this seems to happen at $z\sim 5$ in halos of $\sim 10^{11.5-11.8}~M_{\odot}$, progenitors of $10^{14}~M_{\odot}$ halos at $z=0$. Alternatively, we argue for a possible evidence of a higher SFE in the more massive ALPINE galaxies that is also necessary for the quick stellar mass build-up of these galaxies. To conclude, certainly the three effects, namely outflows, halt of gas supplying, and over-efficient star formation, jointly contribute to the $f_{\rm molgas}$ plateau observed for the massive galaxies at $=4.4-5.9$.
\end{itemize}

%
%-----------------------------------------------------------------------

\begin{acknowledgements}
This paper is based on data obtained with the ALMA Observatory, under Large Program 2017.1.00428.L. ALMA is a partnership of ESO (representing its member states), NSF(USA) and NINS (Japan), together with NRC (Canada), MOST and ASIAA (Taiwan), and KASI (Republic of Korea), in cooperation with the Republic of Chile. The Joint ALMA Observatory is operated by ESO, AUI/NRAO and NAOJ. 
This program receives financial support from the French CNRS-INSU Programme National Cosmologie et Galaxies. 
F.P., F.L., A.C., C.G., and M.T. acknowledge the support from grant PRIN MIUR 2017 - 20173ML3WW$\_$001.
S.F. is supported by the Cosmic Dawn Center of Excellence funded by the Danish National Research Foundation under the grant No.~140.
J.D.S is supported by the JSPS KAKENHI Grant Number JP18H04346, and the World Premier International Research Center Initiative (WPI Initiative), MEXT, Japan.
G.C.J. acknowledges ERC Advanced Grant 695671 ``QUENCH'' and support by the
Science and Technology Facilities Council (STFC). 
M.B. acknowledges FONDECYT regular grant 1170618. 
E.I. acknowledges partial support from FONDECYT through grant No.~1171710.
S.T. acknowledges support from the European Research Council (ERC) Consolidator Grant funding scheme (project “ConTExt”, grant number: 648179). The Cosmic Dawn Center (DAWN) is funded by the Danish National Research Foundation under grant No.~140.
L.V. acknowledges funding from the European Union’s Horizon 2020 research and innovation program under the Marie Sklodowska-Curie Grant agreement No.~746119.
\end{acknowledgements}


\end{document}